\documentclass[12pt]{article}
\usepackage{latexsym,amssymb,amsmath,slashed,cite}
\usepackage{float}
\usepackage[dvips]{graphicx}
\usepackage{comment}
\usepackage{color}
\usepackage{hyperref}
\hypersetup{colorlinks=true,urlcolor=blue,linkcolor=black,citecolor=blue}

\def\rmi{{\rm i}}

\newcommand{\bbox}{\lower.2ex\hbox{$\Box$}}

\csname @addtoreset\endcsname{equation}{section}
\usepackage{ifpdf}
\ifx\pdfoutput\undefined
   \pdffalse
   \usepackage{cite}
 \else
   \pdfoutput=1
   \pdftrue
\newcommand{\rf}[1]{(\ref{#1})}
\usepackage{color}

\def\be{\begin{equation}}
\def\ee{\end{equation}}
\newcommand{\ba}{\begin{eqnarray}}
\newcommand{\ea}{\end{eqnarray}}

\newcommand{\lp}{\left(}
\newcommand{\rp}{\right)}
\newcommand{\ls}{\left[}
\newcommand{\rs}{\right]}

\newcommand{\K}{\mathcal{K}}

\newcommand{\vp}{\varphi}

\def\rmi{{\rm i}}

\newcommand{\cN}{{\cal N}}

\newcommand{\N}{{\cal N}}

\def\K{K{\"a}hler}

\def\K{K{\"a}hler}
\def\Mobius{M\"{o}bius~}

\begin{document}

\begin{titlepage}
\hskip 0.5cm
\vskip 0.5cm
\begin{center}
\baselineskip=20pt


{\LARGE {\bf  Maximal Supersymmetry and B-Mode Targets   }}

\

\

\

{ \bf Renata Kallosh$^{1}$, Andrei Linde$^{1}$, Timm Wrase$^2$, Yusuke Yamada$^1$}

\vskip 0.8cm
{\small\sl\noindent
$^1$ SITP and Department of Physics, Stanford University, Stanford, California 94305, USA \\\smallskip
$^2$ Institute for Theoretical Physics,  TU Wien,  A-1040 Vienna, Austria
}


\vskip 2cm

{\bf Abstract}

\end{center}

{\small

Extending the work of Ferrara and one of the authors \cite{Ferrara:2016fwe}, we present dynamical cosmological models of $\alpha$-attractors with plateau potentials for $3\alpha=1,2,3,4,5,6,7$. These models are motivated by geometric properties of maximally supersymmetric theories:  M-theory, superstring theory,  and maximal $N = 8$  supergravity. 
After a consistent truncation of maximal  to minimal supersymmetry in a seven-disk geometry, we perform a two-step procedure: 1)  we introduce a superpotential,  which  stabilizes  the moduli of the seven-disk geometry in a supersymmetric minimum,  2) we add a cosmological sector with a nilpotent stabilizer, which breaks supersymmetry spontaneously and leads to a desirable class of cosmological attractor models. 
These models with $n_s$ consistent with observational data, and with tensor-to-scalar ratio $r \approx 10^{-2}- 10^{-3}$, provide natural targets for future B-mode searches. We relate the issue of stability of inflationary trajectories in these models  to tessellations of a hyperbolic geometry.

}\vspace{2mm} \vfill \hrule width 3.cm \vspace{1mm}
{\footnotesize \noindent kallosh@stanford.edu, alinde@stanford.edu, timm.wrase@tuwien.ac.at, yusukeyy@stanford.edu}
\end{titlepage}
\addtocounter{page}{1}
\setcounter{tocdepth}{2}
\tableofcontents{}
\newpage

\bibliography{supergravity}
\bibliographystyle{toinemcite}

\parskip 9pt 

\section{Introduction}
We would like to specify targets for the future B-mode detectors for the ratio of the tensor to scalar fluctuations, $r={A_t\over A_s}$. We propose models of inflation with definite values of $r$ which can be validated/falsified either by the detection of primordial gravity waves, or by the improved bounds on $r$. At present, the bound is considered to be $r\leq 7\times 10^{-2}$ \cite{Ade:2015xua} at 95\% confidence level. There are many inflationary models consistent with this bound, see for example, \cite{Abazajian:2016yjj,Calabrese:2016eii,Finelli:2016cyd} where the future CMB observations are described. 

Here we will focus on inflationary $\alpha$-attractor models \cite{Kallosh:2013hoa,Ferrara:2013rsa,Kallosh:2013yoa,Galante:2014ifa,Kallosh:2015zsa,Carrasco:2015uma,Carrasco:2015rva}, based on the hyperbolic geometry of a Poincar\'e disk.  Such a disk is beautifully represented by Escher's picture Circle Limit IV with radius squared  $R^2_{\rm E} =3\alpha$. 
The hyperbolic geometry of the disk has the following line element 
\be
ds^2 = {dx^2 +dy^2 \over \Big (1- {x^2+y^2\over 3 \alpha}\Big )^2}\,,
\label{disk}\ee
which describes a disk with a boundary so that $x^2+y^2 <3\alpha$. More details on this are given in Sec. \ref{sec:plateau}.
The original  derivation of  this class of models was based on  superconformal symmetry and its breaking. 

At present one can view  $\alpha$-attractor cosmological models with a plateau potential  as providing a simple explanation, due to the hyperbolic geometry of the moduli space, of the equation relating the tilt of the spectrum $n_s$ to the number of e-folding of inflation $N$:
\be
n_{s} \approx 1-{2\over N}\, \, .
\label{data1}\ee
This equation is valid in the approximation of a large number of e-foldings $N$ and it is in a good agreement with the data. In addition to providing an equation for $n_s$, the hyperbolic geometry leads to the B-mode prediction for  $r={A_t\over A_s}$: 
\be
 r \approx  3 \alpha \, {4\over N^{2}}= R^2_{\rm E} \, {4\over N^{2}}\, ,
\label{data2}\ee
for reasonable choices of the potential and for $\alpha$ not far from $1$. General expressions for $n_{s}$ and $r$ with their full dependence on $N$ and on $ \alpha$ are also known for a large class of models \cite{Kallosh:2013yoa}. They were derived in the slow roll approximation and  they are more complicated than expressions in \rf{data1}, \rf{data2}. Both CMB observables $n_s$ and $r$ follow from the choice of the geometry  and are not very sensitive to the changes in the potential due to attractor properties of these models.

The experimental value of the scalar tilt suggests that $n_{s} \approx 1-{p\over N}$ with $p=2$ is a good fit to the data. Here $p$ controls the order of the pole in the kinetic term of the inflaton and $p=2$ corresponds to a second order pole, see \eqref{disk}. Various considerations leading to this kind of relation between $n_s$ and $N$ were suggested in the past. For example, in \cite{Mukhanov:2013tua}, using the equation of state analysis, it was argued that robust inflationary predictions can be defined by two constants of order one, $p$ and $q$,  so that at large $N$,  $n_s= 1- {p \over N}$ and $r=  24{q\over N^p}$.  We explain in  Sec. \ref{sec:poleinflation} why in hyperbolic geometry $p=2$ and $6 q = R^2_{\rm E}$. Related ideas were developed  in  \cite{Roest:2013fha,Garcia-Bellido:2014gna,Creminelli:2014nqa}.  Specific examples of such models with plateau potentials and $\alpha=1$ include the Starobinsky model \cite{Starobinsky:1980te}, Higgs inflation \cite{Salopek:1988qh} and conformal inflation models \cite{Kallosh:2013hoa}.

The $\alpha$-attractor models  in $\N=1$ supergravity, starting with \cite{Kallosh:2013yoa}, may have any value of $\alpha$ and, therefore any value of $r$. An example of an $\N=1$ supergravity model with a very low level of $3 \alpha=1/3$ is known   \cite{Goncharov:1983mw}, which is actually the very first supergravity model of chaotic inflation.

The general  class of $\alpha$-attractor models \cite{Kallosh:2013hoa,Ferrara:2013rsa,Kallosh:2013yoa,Galante:2014ifa,Carrasco:2015uma,Kallosh:2015zsa,Carrasco:2015rva},  can be related to string theory in the following sense: the  effective supergravity model is based on two superfields, one is the inflaton, the other one is often called  a stabilizer. It is a nilpotent superfield  which is present on the D3 brane \cite{Ferrara:2014kva,Kallosh:2014wsa}. 

When the geometry of these models is associated with half-maximal $\cN=4$ supergravity \cite{Cremmer:1977tt} and the maximal $\N=4$ superconformal theory  \cite{Bergshoeff:1980is}, one finds   \cite{Kallosh:2015zsa} that 
the lowest value of $3\alpha$ in these models is $1$. It corresponds to a {\it unit size Escher  disk with }
 \be 
R^2_{\rm E} =3\alpha=1\,.
\ee

Note that the relevant value of $r$ is three times smaller than that of the Starobinsky model, Higgs inflation model and conformal inflation models, corresponding to $\alpha=1$,    and provides a well motivated  B-mode target $r\sim 10^{-3}$,  as explained in \cite{Kallosh:2015zsa}. 

In \cite{Ferrara:2016fwe} it was shown that starting with gravitational theories of maximal  supersymmetry, M-theory,  string theory and maximal supergravity,  one finds  a  seven-disk manifold defined by seven complex scalars.  Two assumptions were made in  \cite{Ferrara:2016fwe}:

The first assumption was  that  there exists a dynamical mechanism  which realizes some conditions on these scalars, given in eq.  (4.17) in \cite{Ferrara:2016fwe}. 
If these conditions can be realized dynamically, it would mean that in each case the kinetic term of a single remaining complex scalar  is defined by a hyperbolic geometry with 
\be
R^2_{\rm E}= 3\alpha= 1,2,3,4,5,6,7\,.
\label{discrete}\ee
 
 The second assumption in  \cite{Ferrara:2016fwe}  was that 
 these models can be developed further to produce the inflationary potential of the $\alpha$-attractor models with a plateau potential in a way consistent with the constraints.  This would make the models proposed in  \cite{Ferrara:2016fwe}  legitimate cosmological models,  with specific predictions for $n_s$  in \rf{data1} and $ r$ in \rf{data2}, \rf{discrete}.
 
The purpose of this paper is to show how to construct such dynamical models, thereby validating the assumptions made in \cite{Ferrara:2016fwe}. This will explain how starting from from the seven-disk manifolds of maximal $\cN=8$ supersymmetry models to derive the minimal $\cN=1$ supersymmetric cosmological models with B-mode targets scanning the region of $r$ between $10^{-2}$ and $10^{-3}$. 

\section{Capturing infinity in a finite space: plateau potentials}\label{sec:plateau}
\begin{figure}[h!]
\begin{center}
\includegraphics[width=6cm]{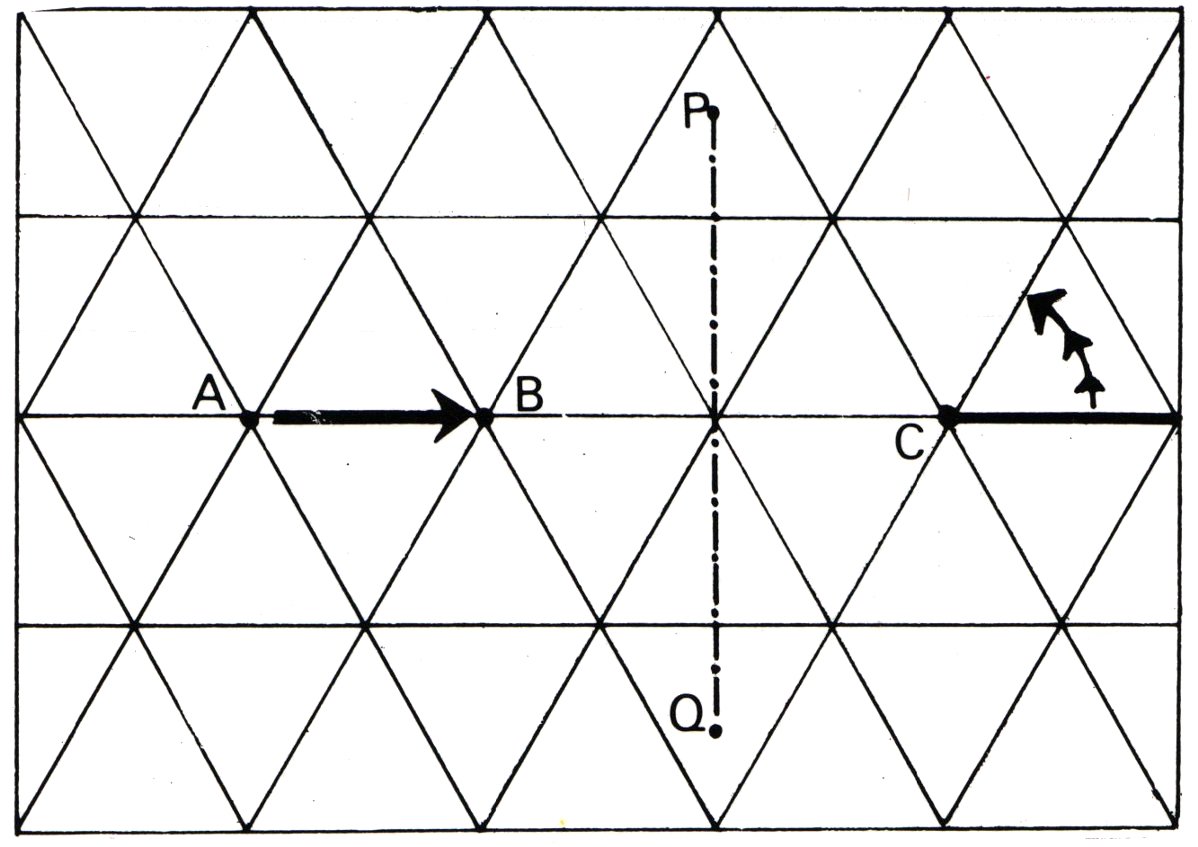}
\caption{\it A tessellation of the flat surface plane  covered with equilateral triangles. The shift of the whole plane over the distance\,  AB,  will cover the underlying pattern  again. This is a\,  {\bf translation}\, of the plane. We can also turn the duplicate through 60 degrees about the point\, C, and we see that  it covers the original pattern exactly. This is a\, {\bf rotation}. Also after a\, {\bf reflection}\, in the line\, PQ, the pattern remains the same.}\label{fig1}
\end{center}
\vspace{-0.5cm}
\end{figure}

Escher was inspired by islamic tilings in Alhambra and he produced beautiful art using a tessellation of the flat surface. The line element of it is
\be
ds^2= dX d\bar X = dx^2+ dy^2\, ,  \qquad X=x+\rmi y\,.
\ee
A tessellation is the tiling of a  plane  using one or  more  geometric shapes, called tiles, with no overlaps and no gaps. Consider a simple example when in the plane the whole surface is covered with equilateral triangles, as in  Fig. \ref{fig1}. 

\begin{figure}[h!t!]
\centering
\includegraphics[width=.24\textwidth, angle=90]{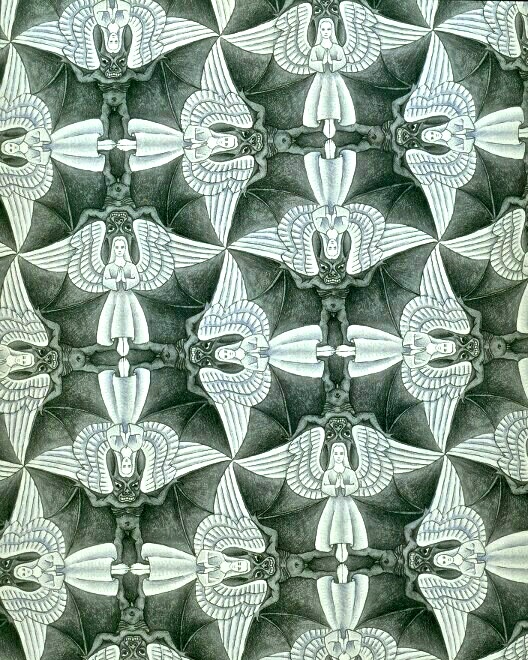}
\hspace{.1\textwidth}
\includegraphics[width=.24\textwidth, angle=90]{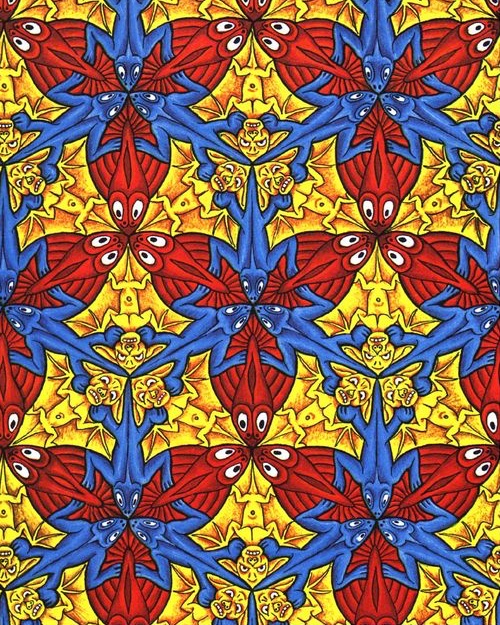}
\caption{\footnotesize  Left: Escher's  tessellation of the flat surface plane  with Angels and Devils design. Right: Escher's tessellations of the flat surface plane  with Lizard/Fish/Bat design.}
\label{fig2}
\end{figure}

The symmetry elements of the tessellation there include {\bf  translations, rotations and reflections}, for example
\be
X\rightarrow X+ a, \quad \bar X \rightarrow \bar X + \bar a \, ,\quad 
X \rightarrow X e^{\rmi\beta}\,, \quad \bar X \rightarrow \bar X e^{-\rmi\beta}\,, \quad 
 X\rightarrow  - X \,,
\ee
and combinations of these.
Escher has reached a perfection in his tessellations of the flat surface, see for example Fig. \ref{fig2} where he had to cut a repeating pattern to fit it into a finite space of the picture. 
For a long time Escher struggled to  produce an infinitely repeating pattern in a finite figure. His desire was {\it to capture infinity in a finite space}. 

\begin{figure}[h!]
\centering
\includegraphics[width=.28\textwidth]{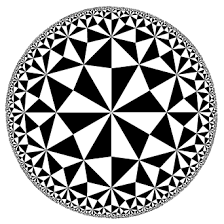}
\hspace{.1\textwidth}
\includegraphics[width=.26\textwidth]{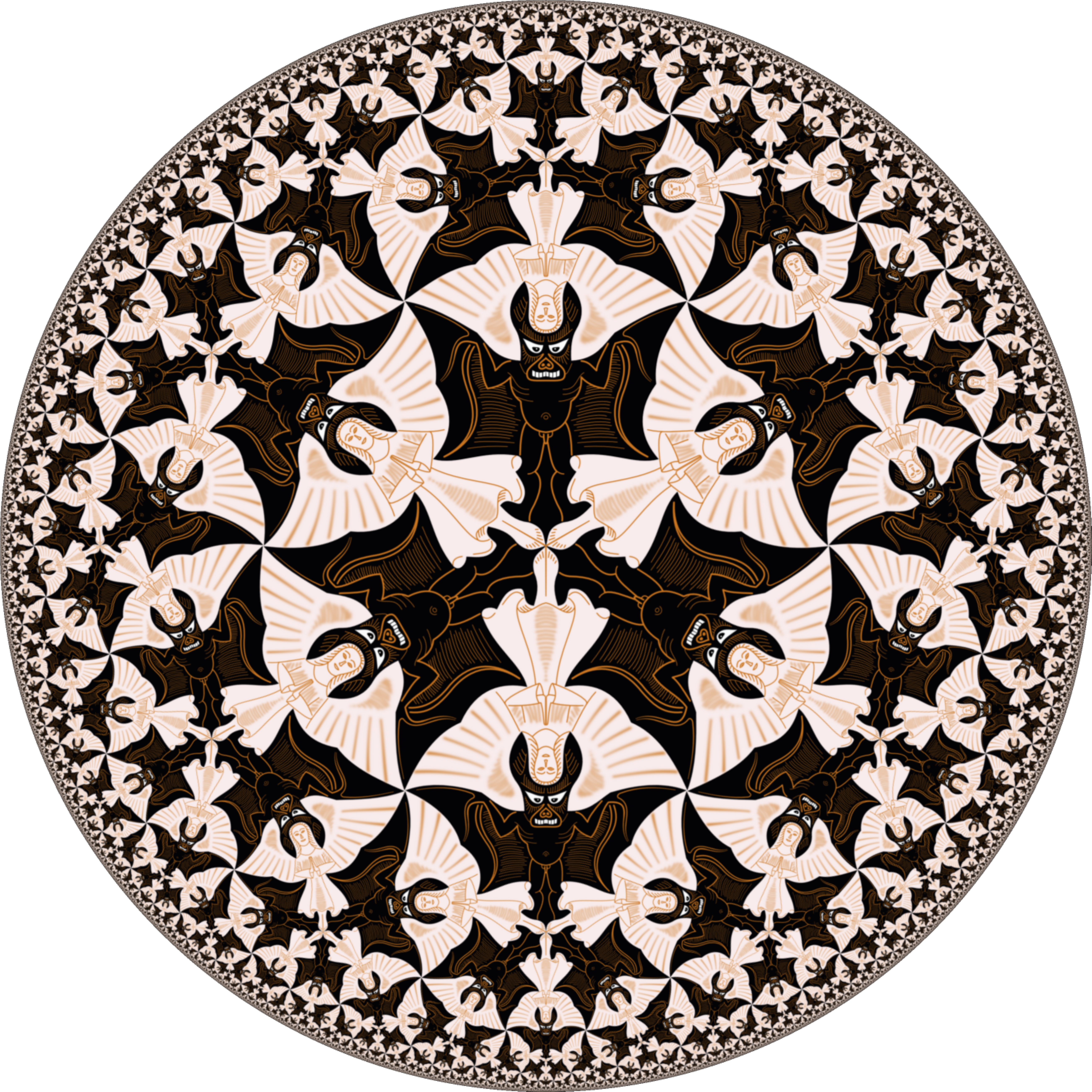}
\caption{\footnotesize  Left: Coxeter's   tessellation of a hyperbolic plane by triangles (Poincar\'e disk model). Right: Escher's  tessellation  for the hyperbolic  tiling for the woodcut  Angels and Devils, ``Circle Limit IV''.}
\label{fig3}
\end{figure}

It was Coxeter who  gave Escher the idea for the Poincar\'e disk. 
When Escher saw the figure of the tessellation of the hyperbolic plane by triangles produced by Coxeter  in \cite{Cox}, see Fig. \ref{fig3}, left, he realized that this  solves his problem. The figure's hyperbolic tiling, with triangular tiles diminishing in size and repeating (theoretically) infinitely within the confines of a circle, was exactly what Escher had been looking for in order to {\it capture infinity in a finite space}.  This allowed him to produce  his well-known Circle Limit  woodcuts, see the Angels and Devils Circle IV,  in Fig. \ref{fig3}, right.

\begin{figure}[h!]
\centering\vspace{-.5cm}
\includegraphics[width=7.5cm]{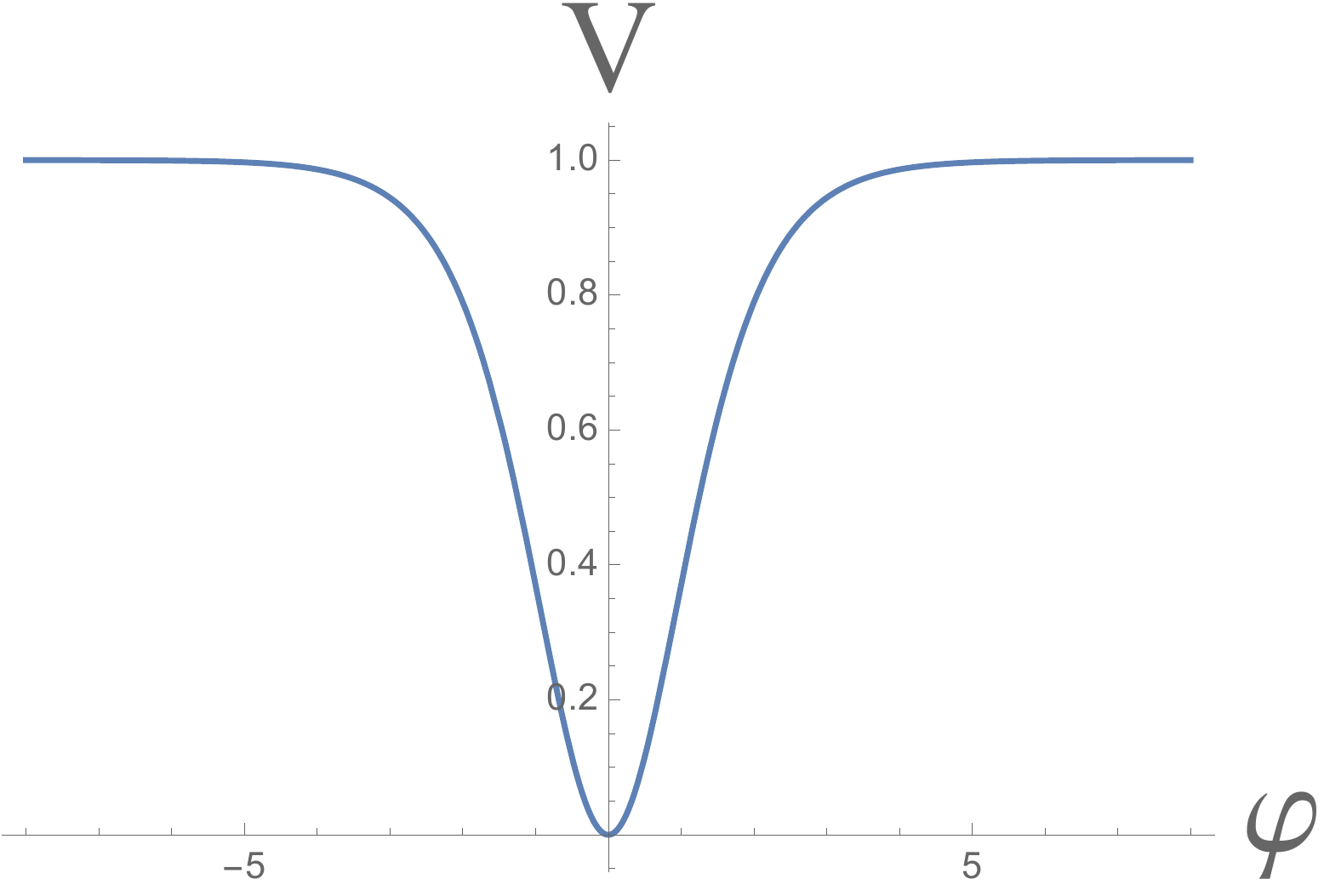}
\caption{A plot of $V(\vp) =\tanh^2 \lp{\vp/\sqrt 2}\rp $. }\label{fig4}
\end{figure}

The complex disk coordinates $Z$  describing the hyperbolic geometry are particularly suitable for providing a mathematical meaning to the concept of  {\it capturing infinity in a finite space}. Namely, a line element of the Poincar\'e disk of radius squared $3\alpha=1$ can be given as \cite{Carrasco:2015rva} 
\be
ds^2 = {dx^2 +dy^2 \over  (1- {x^2-y^2} )^2}=   {dZ d\bar Z \over (1-Z\bar Z )^2}= {1\over 2} \, \,  {d \vp ^2 +d \theta^2\over \cos^2(\sqrt{2} \theta)}\, , \qquad Z= \tanh \Big ({\vp +\rmi\theta\over \sqrt 2}\Big )\,.
\ee
 The angular variable $\theta$ is periodic but the variable $\vp$ is unrestricted.
\be
x^2+y^2 < 1  \qquad \Longleftrightarrow \qquad    \tanh^{2}  \lp\frac{\vp}{\sqrt 2}\rp  <1 \, \qquad -\infty < \vp < +\infty\, , \qquad 0<\theta <2\pi\,.
\ee
We have a map from a finite variable $x^2+y^2 < 1$ describing an inside of the disk to an infinite variable  $-\infty < \vp < +\infty $,  realized by the fact that $\tanh^{2}\lp  {\vp/\sqrt 2}\rp  <1$: the origin of the plateau potential for inflation in Fig. \ref{fig4} can be traced to Escher's concept  of {\it capturing infinity in a finite space}.

We will see that in our cosmological models the angular variable $\theta$ will be quickly stabilized at $\theta=0$ whereas the inflaton field $\vp$ will have a plateau type potential  $V\sim \tanh^{2} \lp {\vp/\sqrt 2}\rp $ in canonical variables, corresponding to a simple potential in the disk variables $V\sim Z\bar Z$.

\begin{figure}[h!]
\centering
\includegraphics[width=.53\textwidth, origin=top]{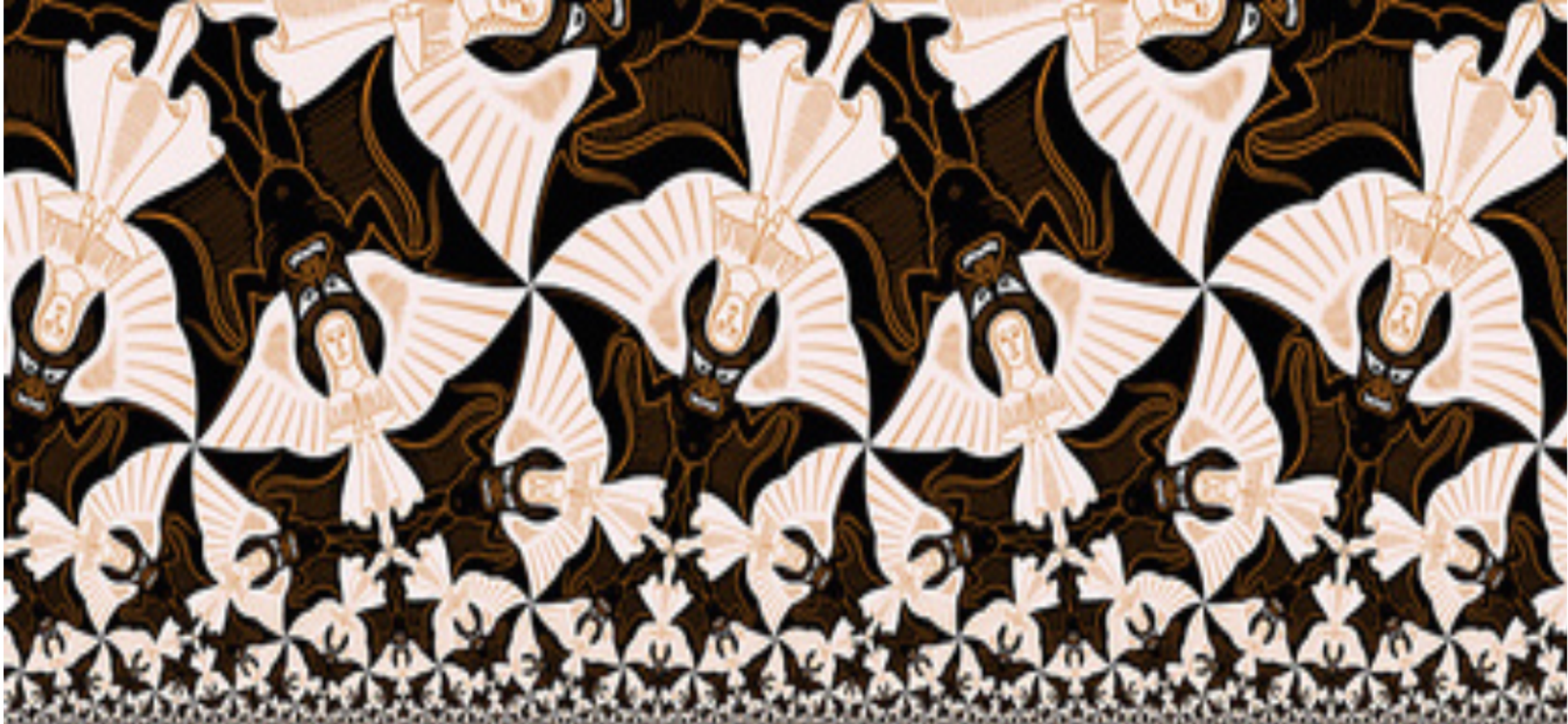}
\hspace{.05\textwidth}
\includegraphics[width=.39\textwidth]{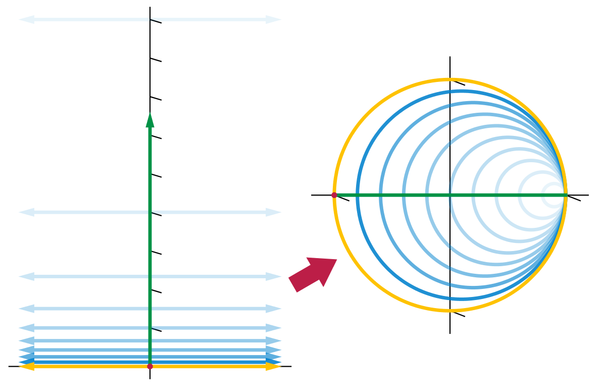}
\caption{\footnotesize  Left: Escher's   tessellation of a hyperbolic half-plane  with ``Angels and Devils" design. Right: Cayley transform of upper complex half-plane to a hyperbolic  disk}
\label{fig5}
\end{figure}

The Cayley transform relates the upper half plane coordinate $X$, ${\rm Im} X > 0$, to the interior of the disk coordinate $Z$, $|Z|<1$. For cosmological models this relation was studied in \cite{Cecotti:2014ipa}:
\be
 X = \tilde x+\rmi \tilde y= \rmi \, {1+Z\over 1-Z}\, ,  \qquad \tilde y>0 \, , \qquad Z= {X-\rmi\over X+\rmi}\, ,  \qquad Z\bar Z<1\, .
\ee
Tessellation of the hyperbolic half-plane are defined by its symmetries, by {\it \Mobius transformations}.  The line element in half-plane variables, see the left part of Fig. \ref{fig5}, with $\tilde y >0$ is
\be
 ds^2_{hp}=  -  {d X d \bar X \over (X-\bar X)^2} =   {d\tilde x^2 + d\tilde y^2\over  4\tilde y^2} \,.
\label{hp} \ee
 It corresponds to the unit size disk geometry
\be
ds^2_{d}=  {dZ d\bar Z \over (1-Z\bar Z )^2}= {dx^2 +dy^2 \over (1- {x^2-y^2} )^2}=  ds^2_{hp} \,.
\ee

The symmetries  in half plane include: {\bf translation  of the real part of $X$, dilatation of the entire plane,   inversion, and reflection of the real part of $X$}:
\ba \label{ax}
&X\rightarrow X+ b\,, \quad \bar X \rightarrow \bar X +  b\,, \\
 \cr  \label{sc}
 & X\rightarrow  a^2 X\,, \\ 
 \cr  \label{in}
 &X \rightarrow  - 1/X\,, \\
 \cr  \label{re}
 & X+\bar X  \rightarrow  - (X+\bar X)\,.
\ea
The first three separate transformations can be also given in the form 
\be
X \rightarrow {a X+b\over cX +d},  \qquad \Delta \equiv ad-bc\neq 0 \,,
\label{sl}\ee
where $a,b,c,d$ are real parameters.
The first one, the shift of the real part,  is the case of $c=0$, $a=d=1$, the second one, rescaling,  is $b=c=0$, $ad=1$, the third one, inversion, is $a=d=0$, $b/c=-1$. 

In the case of the disk with $R^2_{\rm E}=3\alpha$ the metric of the disk and half plane become, respectively 
\be
 ds^2_d=  3 \alpha {dZ d\bar Z \over (1-Z\bar Z )^2}=  {dx^2 +dy^2 \over \Big (1- {x^2+y^2\over 3 \alpha}\Big )^2}\,,
\label{metric}\ee
\be
ds^2_{hp}=  -  3\alpha {d X d \bar X \over (X-\bar X)^2} = {d\tilde x^2 + 3\alpha \, d \tilde y^2 \over  4 \tilde y^2}\,.
\label{pole}
\ee
The curvature of the moduli space associated with the metric in \rf{metric} or \rf{pole} is given by
\be
{\cal R} = - {2\over 3 \alpha}\, . 
\ee
 
\section{Pole inflation: hyperbolic geometry and attractors} \label{sec:poleinflation}
 
In the form \rf{pole} it is particularly clear what is the origin of the $n_s$ and $r$ equations in \rf{data1} and \rf{data2}. For a constant axion $\tilde x$ the hyperbolic geometry line element is
\be
ds^2_{hp}|_{\tilde x=c}=    3\alpha\,{  d \tilde y^2\,  \over \tilde y^2} \,.
\label{attr}\ee
It was explained  in \cite{Galante:2014ifa} that in general, if one starts with a kinetic term for scalars in the form
\be
\mathcal{L}_{kin} = { a_p\over 2}\,  { d  \rho^2 \over    \rho^p} \,,
\label{pPole}\ee
and assumes that inflation takes place near $\rho=0$, so that the potential is 
\be
V\sim V_0 (1- c\rho+ \dots)\,,\qquad c>0\,,
\label{V} \ee
then one finds that at large $N$ (assuming $p>1$)
\be
n_s= 1- {p\over p-1} {1\over N}\, , \qquad r= {8 c ^{p-2\over p-1} \, a_p ^{1\over p-1} \over (p-1)^{p\over p-1} } {1\over N^{p\over p-1}}\,.
\label{p}\ee
Note the following features of the general pole inflation models\cite{Galante:2014ifa} described above
\begin{itemize}
\item For $p=2$  the model displays an attractor behavior, where the dependence on $c$ in the potential \rf{V} is absent (without absorbing such a dependence into a redefinition of the residue of the pole as studied in \cite{Broy:2015qna}\footnote{In the  models of pole inflation in \cite{Broy:2015qna} a slightly different framework was proposed. It was suggested to set $c=1$ via a rescaling of $\rho$ followed by a change of the kinetic term $a_p \rightarrow \tilde{a}_p=c^{2-p} a_p$.}). In such a case \rf{p} simplifies to
\be
n_s= 1-  {2\over N}\, , \qquad r=  a_p  {8\over N^2}\,.
\ee
This case is realized in hyperbolic geometry with $ 2 a_p= R^2_{Es}$. It is interesting that the attractor features of cosmological models starting with the hyperbolic geometry follow from one of the symmetries shown in \rf{sc}. Namely, the kinetic term in \rf{pPole} for $p=2$ is invariant under the change of the parameter $c$ in the potential since if and only if $p=2$ we have for $c \rho = \rho'$ that ${ d  \rho^2 \over    \rho^2} = { d ( \rho')^2 \over (\rho')^2}$. The corresponding symmetry of the hyperbolic half plane is the dilatation.
\item This dilatation of the half plane in eq. \rf{sc}, leading to the attractor property of this particular pole inflation, will be shown below to also lead to the stabilization of the inflationary trajectory, once this symmetry as well as the inversion symmetry in \rf{in} are implemented as symmetries of the \K\, potential.
\end{itemize}

\section{Tessellation,  \K\,  frame,  stability }
The potential in $\cN=1$ supergravity depends on a \K\, potential $K$ and a superpotential $W$
\be
V= e^K (|DW|^2 - 3 |W|^2)\,.
\ee
Here we explain the choices of the \K\, frame, following \cite{Carrasco:2015uma},  emphasizing the relation to the elements of the tessellation and the stability of the inflaton directions.
In  \cite{Ferrara:2016fwe}   the standard  form of the  {\K} potential  was used describing a unit size Poincar\'e disk
\be
K_O=  - \ln(-\rmi(\tau -\bar \tau)) \ .
\label{HPold}\ee
The corresponding line element/kinetic term for the scalar 
\be
ds^2={ d \tau d \bar \tau\over (\tau-\bar \tau)^2}
\label{kintau}\ee
has a M\"obius  symmetry
  \be
\tau'= {a\tau+b\over c\tau +d}\,,  \qquad ad-bc\neq 0 \ ,
\label{sl2}\ee
where $a,b,c,d$ are real parameters. The axion-dilaton pair, using notation in \cite{Carrasco:2015uma},  is for $3\alpha=1$
\be
\tau = \chi - \rmi e^{-\sqrt{2}\vp}\,.
\label{axiondilaton}\ee
We are not in space-time anymore but in the moduli space of scalars fields, therefore instead of $X= \tilde x +\rmi \tilde y$ in eq. \rf{hp} we are using the holomorphic variable $\tau(x)$, where $x \equiv x^\mu$ denotes the space-time dependence of $\tau$.
 
Note that the tessallation of the hyperbolic plane in $\tau$ variables consists of a few independent operations (subgroups of the \Mobius group). The hyperbolic line element  in \rf{hp}, \rf{kintau} is invariant under all transformations of this group. In terms of the axion field $\chi(x)$ and the scalar field $\vp(x)$ in \rf{axiondilaton} the relevant symmetries in \rf{ax} are a {\it shift of the axion by a constant} and a {\it reflection of the axion} in \rf{re}
\ba
\chi(x) &\rightarrow& \chi(x) + b \,,
\label{axi} \\
\chi(x) &\rightarrow&  - \chi(x)\,.  
\label{ref} 
\ea
Let us look at the relevant tessellations of the half plane in Fig. 5 left and compare it with the tessellations in Fig. 1 of the flat full plane. These two symmetries in \rf{axi} and \rf{ref} are the obvious ones.  According to  \rf{axi} one can shift the Angels and Devils to the right (for positive $b$) or to the left  (for negative $b$), the same way as it is done in Fig. 1 of the flat full plane. There the whole plane, not only an upper half of it, is shifted to the right by the distance AB, and the pattern covers the plane again.  According to  \rf{ref} we can choose a vertical line, like in Fig. 1 it is a line PQ, and we can make a reflection in this line, preserving the pattern, see also Fig. 2 left, which shows the Angels and Devils away from the boundary of a half plane, where one can see clearly the existence of such PQ lines. Altogether, this is a convincing argument to give the symmetries of the hyperbolic plane in \rf{axi} and \rf{ref} a name: {\it Tessellation Set 1}.

The other two symmetries of the geometry in \rf{kintau} are the {\it inversion} and the {\it scaling} 
\ba
\tau &\rightarrow& - 1/\tau\,,
\label{inv}\\
\tau &\rightarrow&  a^2 \tau\,.
\label{sca}
\ea 
These two symmetries are absent in a full plane, see for example Fig. 2 left. However, in the half plane which has a boundary, these symmetries control the fact that near the boundary the Angels and Devils in Fig. 5 left are getting smaller, still preserving the pattern. These are rather non-trivial tessellations inherited from the finite size hyperbolic disk tessellations in Fig. 3. These are symmetries responsible for capturing infinity in a finite space. We will give the symmetries of the hyperbolic plane in \rf{inv} and \rf{sca} the name:  {\it Tessellation Set 2}.

{ \it The \K\, potential in \rf{HPold}  is  invariant under the axion shift symmetry \rf{axi} and axion reflection \rf{ref}, i.e. under Tessellation Set 1. However, it breaks the remaining symmetries of the geometry:  inversion symmetry \rf{inv} and the scaling symmetry \rf{sca}, it is not invariant under  Tessellation Set 2. }

In case when the axion ${\rm Re} \, \tau$ is an inflaton field, and the \K\, potential is of the form 
\be
e^{K}=  {1\over ({\rm Im}\, \tau )^{3\alpha}}\,,
\ee
the potential tends to have a run-away factor depending on the sinflaton, ${\rm Im} \tau$:
\be
V= {1\over ({\rm Im}\, \tau )^{3\alpha}} (|DW|^2 - 3 |W|^2)\,.
\label{V0}\ee
This is known in string theory as the \K\, moduli problem.  In the string theory/supergravity context the KKLT construction \cite{Kachru:2003aw} is one way to stabilize these type of moduli; another one is LVS  \cite{Balasubramanian:2005zx}. In the axion monodromy inflation \cite{McAllister:2014mpa},   this problem was addressed in models without supersymmetry, where the 2-form axion does not have a susy partner.

For  $\alpha$-attractor models a new \K\, frame was proposed in \cite{Carrasco:2015uma} where it was argued that in case that the pseudo-scalar  is a heavy field and the scalar field is a light one, it is relatively easy to stabilize the axion. The corresponding \K\, potential takes the form
\be
K_{\rm new}=  -{1\over 2}  \ln \Big (-{(\tau -\bar \tau)^2\over 4 \tau \bar \tau}
\Big ) \ .
\label{HPnew}\ee
In terms of symmetries corresponding to a  set of a hyperbolic plane tessellations, it is complimentary to the \K\, potential in \rf{HPold}.

{ \it The \K\, potential in \rf{HPnew}  is  invariant under the inversion symmetry \rf{inv} and the scaling symmetry \rf{sca}, i.e. under Tessellation Set 2.
 However, it breaks the remaining  symmetries of the geometry:  the axion shift symmetry \rf{axi} and the axion reflection symmetry \rf{ref}. It is not invariant under Tessellation Set 1}. The \K\, potential has a shift symmetry for a dilaton, complemented by a rescaling of the axion  \cite{Carrasco:2015uma}
\be
\chi \rightarrow {a\over d} \chi\,, \qquad \varphi \rightarrow  \varphi + {1\over \sqrt 2} \ln {a\over d}\,.
\ee
At $\chi=0$ one finds that 
\be
e^{K_{\rm new} |_{\tau =-\bar\tau}}= e^{ -{1\over 2}  \ln  (-{(\tau -\bar \tau)^2\over 4 \tau \bar \tau} )  | _{\tau =-\bar \tau} }
=1 \ .
\label{HPnew1}\ee
and there is no  run-away behavior of the potential. Good choices of the superpotentials produce desirable cosmological models with
\be
V_{\rm new} =  |DW_{\rm new}|^2 - 3 |W_{\rm new} |^2\,.
\label{pot}\ee
Thus, in the new frame, the \K\, potential has an inflaton flat direction, which is lifted by the superpotential. 

Clearly, the relation between the old frame \rf{HPold} and the new frame \rf{HPnew} satisfies the \K\, symmetry
\be
K_{\rm new} \rightarrow K - \ln \Phi \bar \Phi\,, \qquad  W_{\rm new} \rightarrow W \cdot \Phi \,,
\ee 
and one theory can be related to the other.
However, the potential in \rf{pot} is expected to be small to describe slow roll inflation, it will produce small deviation of the flatness of the theory in the inflaton direction. Therefore the new choice of the frame \rf{HPnew} has an advantage over the choice in \rf{HPold} with regard to the stabilization of the inflationary trajectory. 
Starting with $K_O$ one has to look for a $W_O$ which will make the inflationary potential approximately flat despite the fact that we started with the strong run-away potential in \rf{V0}. Instead, we start with $K_{\rm new}$ with a flat inflaton potential broken by a small $W_{\rm new}$. 
In technical terms, it was a flip of one set of a hyperbolic plane tessellation in \rf{axi}, \rf{ref}, which we called Tessellation Set 1,   to  another set in hyperbolic plane tessellation in \rf{inv}, \rf{sca}, which we called Tessellation Set 2,  which has created a desirable stabilization effect.

\section{Dynamical stabilization of seven-disk  models }\label{7d}
The seven disk model in \cite{Ferrara:2016fwe} has the following \K\, potential
\be
K = -\sum_{i=1}^7 \log (T_i+\bar T_i)\,.
\ee
Here we use for each of the seven unit size disks the following variables $T_i = \rmi \tau_i$ and $T_i= e^{-\sqrt 2 \vp_i} + \rmi\chi_i$. It was argued in \cite{Ferrara:2016fwe} that such a \K\, potential for the seven disks can be derived from maximally supersymmetric models with 8 Majorana spinors, M-theory, superstring theory, maximal supergravity, by a consistent truncation to a minimal supersymmetric model with a single Majorana spinor.
 
Following  \cite{Carrasco:2015uma} and as explained in the previous section, we make a choice of the {\K} frame where the {\K} potential has an inflaton shift symmetry: we use the following {\K} potential
\be
K_0 = -\frac12 \sum_{i=1}^7 \log   \frac{(T_i+\bar{T}_i)^2}{4T_i \bar{T}_i} \,.
\ee
An equivalent form can be given in disk variables
\be
K_0 = -\frac12 \sum_{i=1}^7 \log   \frac{(1-Z_i\bar Z_i)^2}{(1-Z_i^2)(1-\bar Z_i^2)} \,.
\ee
At this level the seven-disk theory has an unbroken $\cN=1$ supersymmetry and no potential. Our first step is to find a superpotential and scalar potential depending on the disk variables $T_i$ which has an $\cN=1$ supersymmetric minimum producing the constraints on the moduli which were imposed in  \cite{Ferrara:2016fwe}, namely:
\ba 
 &3\alpha =7 \, :  &T_1=T_2=T_3=T_4=T_5=T_6= T_7 \equiv  T  \cr
 &3\alpha =6 \, :  &T_1=T_2=T_3=T_4=T_5=T_6\equiv  T  \, ,  \qquad  T_7={\rm const }
\label{all 6}\cr
 &3\alpha =5 \, :  &T_1=T_2=T_3=T_4=T_5 \equiv  T  \, ,   \qquad   T_6=T_7={\rm const }
\label{all 5}\cr
&3\alpha =4 \, :  &T_1=T_2=T_3=T_4 \equiv  T  \, ,   \qquad   T_5=T_6=T_7={\rm const }
\label{all 4}\cr
&3\alpha =3 \, :  &T_1=T_2=T_3 \equiv  T  \, ,   \qquad   T_4=T_5=T_6=T_7={\rm const }
\label{all 3}\cr
&3\alpha =2 \, :  &T_1=T_2 \equiv  T  \, ,   \qquad   T_3=T_4=T_5 =T_6=T_7={\rm const }
\label{all 2}\cr
&3\alpha =1 \, :  &T_1 \equiv  T  \, ,   \qquad   T_2=T_3=T_4=T_5=T_6=T_7={\rm const }
\label{all}
\ea
 
Let us explain for example  how the single $\alpha$-attractor model with $3\alpha =7$ case is achieved here. The kinetic terms for the seven complex moduli originally is
\be
{\cal L}_{kin}= - \sum_{i=1}^7 {d T_i d \bar T_i\over  (T_i + \bar T_i)^2}.
\label{kin}\ee
The K\"ahler potential of each $T_i$ corresponds to the $\alpha$-attractor with $3\alpha=1$.
When the condition that 
\be
T_1= T_2=T_3=T_4=T_5=T_6=T_7 \equiv T
\label{cond}\ee
is enforced the kinetic term becomes
\be
{\cal L}_{kin}= - 7 {d T d \bar T\over  (T + \bar T)^2}\,.
\label{kin1}\ee
This is an $\alpha$-attractor model with $3\alpha =7$ with regard to the kinetic term. 
In the following, we will show how the above identifications~\eqref{all}, or equivalently, $3\alpha=1,\ldots,7$ are realized by a dynamical mechanism.

\subsection{Step 1: Enforcing an $\cN=1$ supersymmetric minimum}
We would like to dynamically enforce that all seven fields (or a subset thereof, see below) move synchronously during inflation so that $T_i - T_j=0$. This can be done via a superpotential that gives a very large mass to the combinations $T_i - T_j$:
\be\label{W07}
W_0 = M \sum_{1\leq i<j\leq 7} (T_i-T_j)^2\,.
\ee
We find that the minimum with unbroken supersymmetry,
\be
D_i W_0 = 2 M \sum_{j\neq i} (T_i -T_j) - \frac{(T_i-\bar{T}_i)}{2 T_i (T_i+\bar{T}_i)}\ M \sum_{1\leq j<k\leq 7} (T_j-T_k)^2 = 0\,,
\ee
enforces the conditions that $T_i - T_j=0$. Since this implies $W_0=0$, this solution corresponds to a supersymmetric Minkowski critical point of the scalar potential.

Let us go to canonically normalized fields $T_i=e^{-\sqrt{2} u_i}(1+\rmi \sqrt{2} a_i )$, such that for $a_i=0$ we have
\be
{\cal L}_{kin} =- \sum_i K_{0,T_i \bar{T}_i} \partial_\mu T_i \partial^\mu \bar{T}_i = -\frac12 \sum_i \lp\partial_\mu u_i \partial^\mu u_i +\partial_\mu a_i \partial^\mu a_i\rp\,.
\ee
The mass matrix at the critical point is diagonalized by introducing the new coordinates $u=\frac{1}{\sqrt{7}} \sum_{i=1}^7 u_i$, $a=\frac{1}{\sqrt{7}} \sum_{i=1}^7 a_i$, $\tilde{u}_{j}=\frac{1}{\sqrt{(8-j)(7-j)}}((7-j)u_{j} - u_{j+1}-u_{j+2}-\ldots -u_7)$ and $\tilde{a}_{j} =\frac{1}{\sqrt{(8-j)(7-j)}}((7-j)a_{j} - a_{j+1}-a_{j+2}-\ldots -a_7)$, $j=1,\ldots,6$. For the kinetic terms, restricting to the saxions (the axions work in the same way), we have
\be
{\cal L}_{kin} = -  \frac12 \sum_i \partial_\mu u_i \partial^\mu u_i =-  \frac12\Bigl( \partial_\mu u \partial^\mu u +  \sum_{j=1}^6 \partial_\mu \tilde{u}_{j} \partial^\mu \tilde{u}_{j} \Bigr)\,.
\ee
The canonical masses in these new variables at the supersymmetric minimum are then given by
\be\label{eq:7diskmasses}
m^2_u = m^2_a= 0\,,\qquad m_{\tilde{u}_{j}}^2 = m_{\tilde{a}_{j}}^2 =3136 e^{-4 \sqrt{\frac 2 7} u}M^2 \,.
\ee
Thus we have kept the `diagonal' directions massless, while making all the `transverse' directions arbitrarily heavy.

Analogously we can `identify' any number $1\leq n < 7$ of the seven fields and decouple all other fields by making them heavy. We simply choose the superpotential
\be
W_0 = M \Bigl( \sum_{1\leq i<j\leq n} (T_i-T_j)^2 +\sum_{k=n+1}^7 (T_k-c)^2 \Bigr)\,.
\label{W0}\ee
This choice of $W_0$ leads to a dynamical realization of the proposal in \cite{Ferrara:2016fwe} that there is a relation between the moduli of the seven disks. 
 In presence of the superpotential \rf{W0} a set of conditions \rf{all} follows from the conditions for a supersymmetric minimum
\be
D_i W_0 =  0\,.
\ee
Again after going to canonically normalized fields $T_i=e^{-\sqrt{2} u_i}(1+\rmi \sqrt{2} a_i )$, the canonical masses at the minimum for the fields $u=\frac{1}{\sqrt{n}}\sum_{i=1}^n u_i$, $a=\frac{1}{\sqrt{n}}\sum_{i=1}^n a_i$,  $\tilde{u}_{j}=\frac{1}{\sqrt{(n+1-j)(n-j)}}((n-j)u_{j} - u_{j+1}-u_{j+2}-\ldots -u_n)$ and $\tilde{a}_{j} =\frac{1}{\sqrt{(n+1-j)(n-j)}}((n-j)a_{j} - a_{j+1}-a_{j+2}-\ldots -a_n)$, $j=1,\ldots,n-1$ and $u_k$, $a_k$, $k>n$ are given by
\be
m^2_u = m^2_a= 0\,,\qquad m_{\tilde{u}_{j}}^2 = m_{\tilde{a}_{j}}^2 = 64 n^2 e^{-4 \sqrt{\frac 2 n} u} M^2\,,\qquad m_{u_k}^2 = m_{a_k}^2 = 64 c^4 M^2 \,.
\ee
Thus we have kept the `diagonal' directions massless, while making all the `transverse' directions arbitrarily heavy.

\subsection{Step 2: Introducing a cosmological sector}
Until now we have only kept the 7 complex moduli resulting from the consistent reduction of $\cN=8$ supersymmetry (from M-theory, superstring theory and maximal supergravity) to $\cN=1$ theory and by allowing the superpotential $W_0$ to depend on these moduli. We found a supersymmetric Minkowski minimum where the solution requires that an assumption in \cite{Ferrara:2016fwe} given in \rf{all} is realized dynamically as a consistency condition of the supersymmetric minimum for the moduli dependent superpotential in \rf{W0}. 

We would now like to lift one of the flat directions, the $u$ direction, and use it for inflaton. We also need to stabilize its axion partner,  the axion $a$,  and we need to keep all other fields heavy during inflation. At this point we define Step 2 in our model building, as was similarly done before in the KKLT uplifting. In its early version in \cite{Kachru:2003aw} it was suggested that a non-perturbative effect in string theory, the effect of the anti-D3 brane, can be used to uplift the AdS vacuum to a de Sitter vacuum. A more recent version of such a stringy uplift, which became a useful tool in supergravity cosmological model building\footnote{In the supergravity context it was first shown in \cite{Antoniadis:2014oya} that the nilpotent multiplet is useful in cosmological model building.},  is to use a nilpotent supermultiplet $S$, signaling the presence of the uplifting anti-D3 brane in the theory.

To that extent we introduce in our supergravity model a new field $S$ that can be taken to be nilpotent, giving a connection to D-brane physics in string theory,\footnote{The stabilizer $S$ does not necessarily have to be nilpotent. Instead, we may use a usual chiral superfield as stabilizer and add the nilpotent superfield only for SUSY breaking in the vacuum. Even in such a case, the cosmological predictions are not affected, if the SUSY breaking scale is sufficiently smaller than the scale of inflation~\cite{Linde:2016bcz}.} i.e. $S^2=0$. It has a canonical {\K} potential
\be\label{eq:Kaehler}
K_1 = K_0 + S \bar{S}\,.
\ee
We also modify the superpotential so that for the case of $n$ identical fields $T_i$ it takes the form
\be \label{eq:Superpotential}
W_1 = W_0+ S f(T_1 + \ldots +T_n) \,,    \qquad 1\leq n \leq 7\,,
\ee
where $f$ is a so far unspecified function. Here we are only interested in the regime of inflation, not the exit stage, so for simplicity we will put to zero the $S$-independent part of $W$, which may exist in addition to $W_0$ above.
We find that the $T_i$, $1\leq i \leq 7$, critical point equations are satisfied for $S=0$, $T_i =T_j$, $i,j \in\{1,\ldots,n\}$, Re$(T_k)=c$, $k \in \{ n+1,\ldots, 7\}$, Im$(T_i)=$Im$(T_k)=0$, and $f(n \text{Re}(T_1))f'(n \text{Re}(T_1)) = 0$. In this case the F-term for $S$ is simply $D_S W = f(n \text{Re}(T_1)$. 

\subsubsection{\boldmath $3\alpha =7$ case}
For $n=7$ the superpotential is given by
\be
W = M \sum_{1\leq i<j\leq 7} (T_i-T_j)^2+ S f(T_1 + \ldots +T_7) \,.
\ee

The critical point equations $\partial_{T_i} V=0$ are all solved for $T_i =T_j$, $\forall i,j$, Im$(T_i)=0$ and $f(7 \text{Re}(\tau_1)) f'(7 \text{Re}(\tau_1)) = 0$. For such a solution we have  $V = f^2 $, where we have taken $f$ to satisfy $\overline{f(T)} = f(\bar{T})$ so that in particular $\overline{f(7 \text{Re}(T_1))}=f(7 \text{Re}(T_1))$.

We again go to canonically normalized fields as above but now we want to study inflation where $u$ is displaced from its minimum but all the other fields are not. So in particular we keep all $a=a_{1i}=u_{1i}=0$, and displace $u$ from $u_{min}$ which is determined via $f(7 e^{- \sqrt{\frac27}u_{min}})f'(7 e^{- \sqrt{\frac27}u_{min}})=0$. During inflation one finds the following masses for the directions transverse to the inflaton $u$
\ba
m^2_a &=& 28 e^{-2\sqrt{\frac 2 7} u} \left((f')^2-f f''\right)+2 f^2 \,,\cr
m_{u_{j}}^2 &=& 3136 e^{-4 \sqrt{\frac 2 7} u}M^2+4  e^{-\sqrt{\frac 2 7} u}f f'\,, \cr 
m_{a_{j}}^2 &=&3136 e^{-4 \sqrt{\frac 2 7} u}M^2+2 f^2 \,,
\ea
where the argument of $f$, $f'$ and $f''$ is $7 e^{- \sqrt{\frac27}u}$ and the prime denotes the derivative with respect to the argument. We see that during inflation it is not guaranteed that all transverse directions remain stable, however, to leading order in large $M$ the above expressions reduce to equation \eqref{eq:7diskmasses} above. Note that the $\tilde{u}_{j}$ will become exponentially light during inflation when $u \gg 1$ (unless $f$ or $f'$ compensate for this). 

We have explained above near eqs.  \rf{kin}, \rf{cond}, \rf{kin1} how the kinetic term becomes equal to the one with $3\alpha = 7$. Now we would like to see what the situation is with a superpotential and a scalar potential. We find that when \rf{cond} is imposed via $W_0$ in eq. \rf{W07} then the superpotential becomes 
\be
W=  Sf(7\, T).
\ee
We may also use the rescaled single disk variable $T'\equiv 7\, T$. The kinetic term and the superpotential are now given by
\be
{\cal L}_{kin}= - 7\, {\partial_\mu T' \partial^\mu \bar T'\over  (T' + \bar T')^2}\, ,\qquad W=  Sf(T').
\ee
This is an $\alpha$-attractor model with $3\alpha =7$. Note that to bring this model to the standard form of the $\alpha$-attractor model with $3\alpha =7$ we have used again a symmetry of the hyperbolic geometry: Tessellation Set 2, responsible for the attractor feature of our models, which allows us to rescale the half plane coordinate. 

The inflationary regime is at $T'=\bar T' = e^{-\sqrt {2/7} \vp}$ and we find that the inflaton Lagrangian is given by 
\be
\mathcal{L}_{inf} = -{1\over 2} \partial_\mu \vp \partial^\mu \vp - f^2\Big (e^{-\sqrt {2/7} \vp}\Big)\,.
\ee 

\subsubsection{\boldmath $3\alpha =1, \dots, 6$  cases}

Likewise we can study the case where only a subset of the $T_i$ align during inflation. We take again the {\K} potential in equation \eqref{eq:Kaehler} and the superpotential 
\be
W_1 =  W_0 + S f(T_1 + \ldots +T_n) \,.
\ee
The critical point equations $\partial_{T_i} V=0$ are all solved for $T_i =T_j$, $\forall i,j \leq n$, Re$(T_k)=c$, $\forall k>n$, Im$(T_i)=$Im$(T_k)=0$, and $f(n \text{Re}(T_1))f'(n \text{Re}(T_1)) = 0$. For such a solution we  have  $V = f^2 $, where we again took  $f$ to satisfy $\overline{f(T)} = f(\bar{T})$ so that in particular $\overline{f(n \text{Re}(T_1))}=f(n \text{Re}(T_1))$.

We can now again calculate the canonically normalized masses during inflation, i.e. when $u$ is displaced form its minimum but all the other fields are not. The canonically normalized masses of the fields transverse to the inflaton $u$ are given by
\ba\label{eq:genericmasses}
m^2_a&=&4 n e^{-2\sqrt{\frac2n} u} \left((f')^2-f f''\right)+ 2  f^2\,,\cr
m_{\tilde{u}_{j}}^2 &=& 64 n^2 e^{-4 \sqrt{\frac2n} u} M^2+4 e^{-\sqrt{\frac2n} u} f f'\,,\cr
m_{\tilde{a}_{j}}^2 &=& 64 n^2 e^{-4 \sqrt{\frac2n} u} M^2 +2 f^2\,,\cr
m_{u_k}^2 &=&64 c^4 M^2  \,,\cr
m_{a_k}^2 &=& 64 c^4 M^2 + 2 f^2  \,,
\ea
where the argument of $f$, $f'$ and $f''$ is $n e^{- \sqrt{\frac2n}u}$ and the prime denotes the derivative with respect to the argument. So again we see that stability during inflation is dependent on the function $f$ (as well as on $M$). 
Note that all the previous results can be obtained as  restrictions of equation \eqref{eq:genericmasses}. The case with 7 aligned fields follows by simply plugging in $n=7$ and dropping the $u_k$ and $a_k$. The results in the previous subsection follow for $f=0$.

\subsubsection{\boldmath $3\alpha =1, \dots, 7$  in disk variables}
We can likewise analyze this model in disk variables in which case we have the following {\K} and superpotential
\ba
K &=& -\frac12 \sum_{i=1}^7 \log \ls \frac{(1-Z_i \bar{Z}_i)^2}{(1-Z_i^2)(1-\bar{Z}_i^2)}\rs+S\bar S\,,\cr
W &=& M \Bigl(\sum_{1\leq i<j\leq n} (Z_i-Z_j)^2 +\sum_{k=n+1}^7 (Z_k-c)^2 \Bigr) + S f(Z_1+\ldots+Z_n)\,.
\ea
Again after going to canonically normalized fields $Z_i=\tanh\lp \frac{\vp_i + \rmi \theta_i}{\sqrt{2}} \rp$, we can calculate the canonical masses for the fields $\vp=\frac{1}{\sqrt{n}}\sum_{i=1}^n \vp_i$, $\theta=\frac{1}{\sqrt{n}}\sum_{i=1}^n \theta_i$, $\tilde{\vp}_{j}=\frac{1}{\sqrt{(n+1-j)(n-j)}}((n-j)\vp_{j} - \vp_{j+1}-\vp_{j+2}-\ldots -\vp_n)$ and $\tilde{\theta}_{j} =\frac{1}{\sqrt{(n+1-j)(n-j)}}((n-j)\theta_{j} - \theta_{j+1}-\theta_{j+2}-\ldots -\theta_n)$, $j=1,\ldots,n-1$ and $\vp_k$, $\theta_k$, $k>n$. We find during inflation, i.e. for arbitrary $\vp$ but $\theta=\tilde{\vp}_{j}=\tilde{\theta}_{j}=\vp_k=\theta_k=0$, that the masses for the fields transverse to the inflaton $\vp$ are given by
\ba\label{eq:genericmassesdisk}
m^2_a&=&\frac{1}{\cosh^4  \frac{\vp}{\sqrt{ 2n}} }\Bigl(  n \left((f')^2-f f''\right) + f f' \sinh  {\sqrt  2\vp\over n}  \Bigr)+ 2  f^2\,,\cr
m_{\tilde{\vp}_{j}}^2 &=& \frac{4 n^2 M^2}{ \cosh^{8} \frac{\vp}{\sqrt{2n}} } - 2 f f'\frac{\tanh\ \frac{\vp}{\sqrt{2n}} }{ \cosh^{2} \frac{\vp}{\sqrt{2n}}  }\,,\cr
m_{\tilde{a}_{j}}^2 &=&\frac{4 n^2 M^2}{ \cosh^{8} \frac{\vp}{\sqrt{2n}} }  +2 f f'\frac{\tanh \frac{\vp}{\sqrt{2n}} }{ \cosh^{2} \frac{\vp}{\sqrt{2n}} } +2 f^2 \,,\cr
m_{\vp_k}^2 &=& 4 (c^2-1)^4 M^2  \,,\cr
m_{a_k}^2 &=& 4 (c^2-1)^4 M^2 + 2 f^2  \,,
\ea
where the argument of $f$, $f'$ and $f''$ is $n \tanh {\vp\over\sqrt{2n}} $ and the prime denotes a derivative with respect to the argument. So as before, stability during inflation depends on the function $f(n \tanh {\vp\over\sqrt{2n}} )$ and $M$. We also see that again the masses of the $\tilde{\vp}_{j}$ fields will become exponentially small for large $\vp$ (unless we choose a very special function $f$). The above equations are valid for $1\leq n \leq 7$ and arbitrary functions $f$.

\section{Examples: Single-disk and two-disk manifolds}

The dynamics of the 14-dimensional moduli space of the seven-disk manifold is complicated. We described it above paying attention to the mass of the non-inflaton moduli and trying to show that there is a possibility for these masses to be positive during inflation, i.e. the unwanted moduli are stabilized in our seven-disk  models. However, it is important to  relate our new results to the well known   single-disc models, as well as to explain the  basic mechanism of the merger of several cosmological attractors in a toy model with only a two-disk manifold, where we can perform a more detailed analysis of the dynamics of the system. 

\subsection{Basic single-disk models}
\begin{figure}[H]
 \vspace*{3mm}
\centering
\includegraphics[width=8cm]{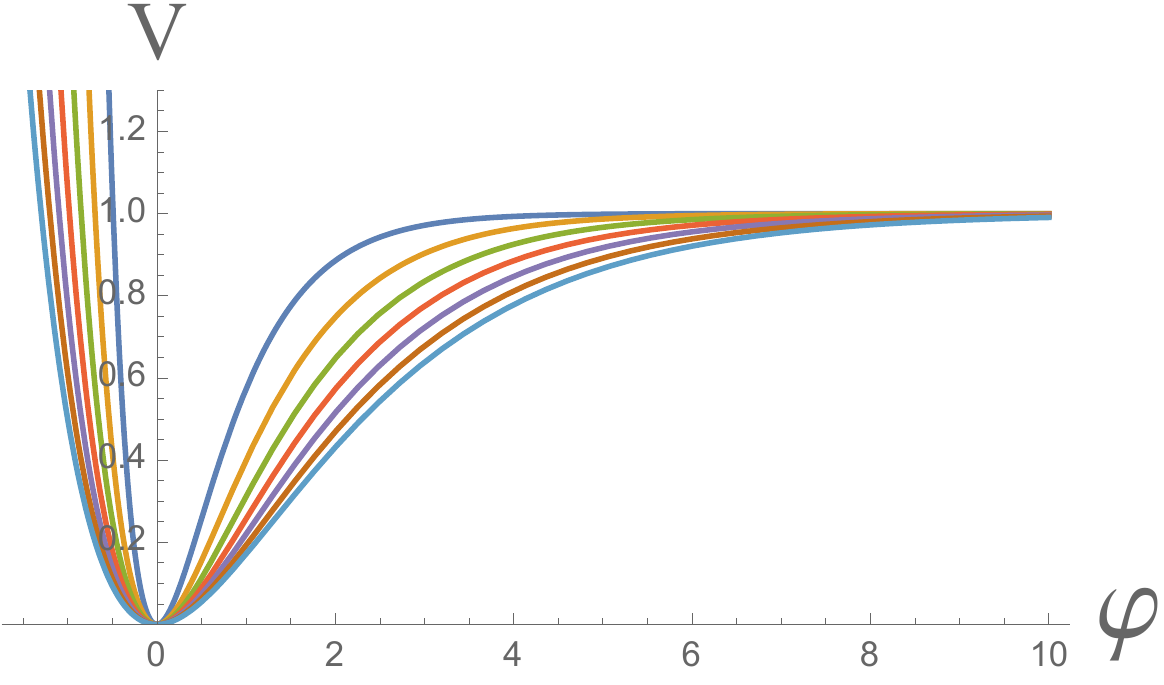}
\caption{\small E-model potentials for  $3\alpha = 1,2, 3,...,7$. The central (blue) line corresponds to  $3\alpha = 1$. The third line from the center (green) corresponds to the supergravity generalization of the Starobinsky model with $\alpha = 1$ \cite{Ferrara:2013rsa,Kallosh:2013xya}.  The outer line shows the potential with $3\alpha = 7$. The value of the inflaton field $\vp$ is shown   in Planck units $M_{p}= 1$; the height of the potential is shown in units of $m^{2}$. }
\label{Emod}
\end{figure}

We begin with models in half-plane variables with
\be\label{K11}
K = -\frac{3\alpha}2  \log   \frac{ (T+\bar{T})^2}{4T \bar{T}}     + S\bar S\,,
\ee
 and with the simple superpotential 
 \be\label{w11}
W =   m S (1-T)  \,.     
\ee
It is convenient to switch to the new variables $T=e^{-\sqrt{2\over3\alpha}\vp}(1+\rmi \sqrt{2} a )$. During inflation, the field $a$ is stabilized at $a = 0$, whereas the field $\vp$ is the canonically normalized  inflaton field with the plateau potential 
\be\label{EEE}
V =  m^{2} \left(1-e^{-\sqrt{2\over3\alpha}\vp}\right)^{2}\,.
\ee
This is the simplest representative of  E-models introduced in \cite{Ferrara:2013rsa,Kallosh:2013yoa,Kallosh:2015zsa}. The theory of initial conditions for inflation in similar theories of  inflation with  plateau potentials was developed in \cite{Linde:2004nz,Carrasco:2015rva}.  The amplitude of the scalar perturbations in such models matches the Planck normalization for  $m \approx  10^{{-5}} \sqrt {\alpha}$ \cite{Kallosh:2015lwa}.  The set of the E-model potentials for $3\alpha = 1,2, 3,...,7$ is shown in Fig. \ref{Emod}.

In what follows, it will be also important for us to know 
the value of  the inflaton field corresponding to the moment when the remaining number of e-foldings of inflation becomes equal to some number $N$. By solving field equations in the leading approximation in $1/N$, one finds \cite{Eshaghi:2016kne}
\be\label{efold}
\vp_N\approx \sqrt {3\alpha \over 2} \log {4 N\over 3 \alpha} \ .
\ee
 For  $\alpha$-attractors  with a more general class of potentials $V =  m^{2}\bigl(1-e^{-\sqrt{2\over3\alpha}\vp}\bigr)^{2n}$ one  has 
\be\label{efold2}
\vp_N\approx \sqrt {3\alpha \over 2} \log {4Nn \over 3 \alpha} \ .
\ee

\begin{figure}[H]
 \vspace*{3mm}
\centering
\includegraphics[width=9cm]{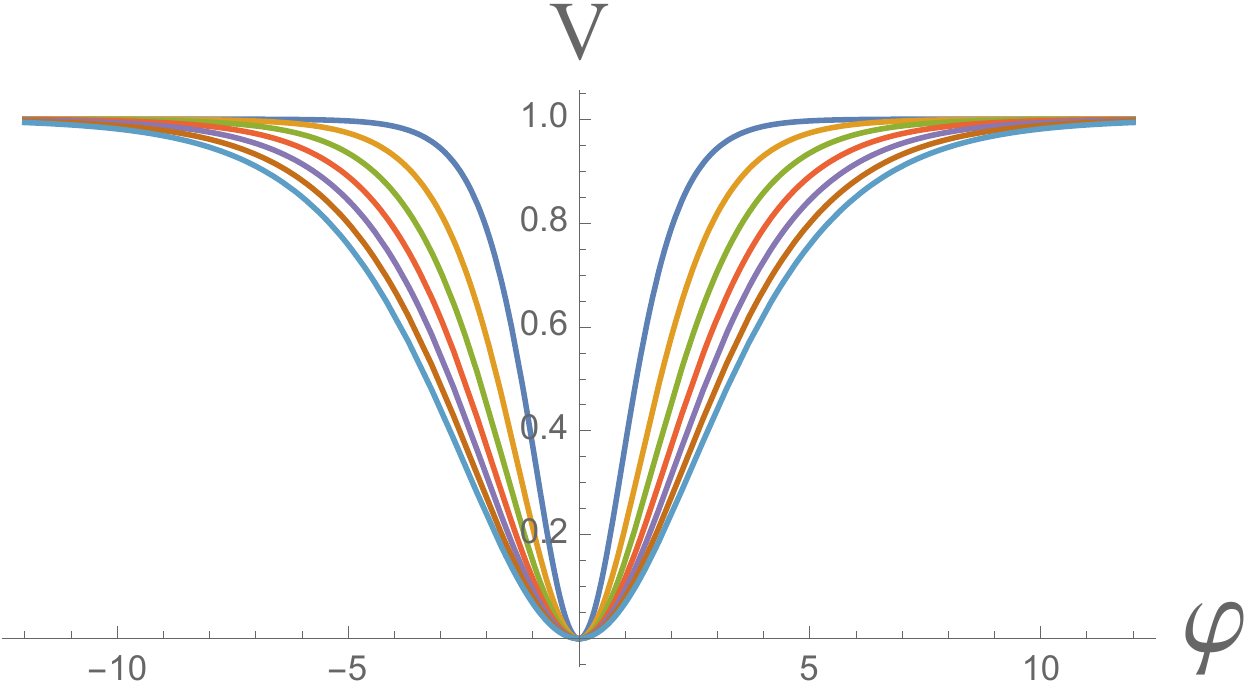}
\caption{\small T-model potentials for  $3\alpha = 1,2, 3,...,7$. The central (blue) line corresponds to the disk with  $3\alpha = 1$. The third line from the center (green) corresponds to the conformal inflation model with $\alpha = 1$ \cite{Kallosh:2013hoa}.  The outer line shows the potential with $3\alpha = 7$. The value of the inflaton field $\vp$ is shown   in Planck units $M_{p}= 1$; the height of the potential is shown in units of $m^{2}$. }
\label{Tmod}
\end{figure}

Now we will consider $\alpha$-attractors in disk variables. The simplest model is described by   
\be\label{DiskK}
K = -\frac{3\alpha}2 \log  \frac{(1-Z\bar Z)^2}{(1-Z^{2})(1-\bar Z^2)} +S\bar S\,, \qquad W  =   m\, S\, Z   \, . 
\ee
This leads to the T-model potential of the inflaton field \cite{Kallosh:2013yoa,Kallosh:2015zsa}
\be
V=   m^{2} \tanh^{2} {\vp\over \sqrt 6\alpha} \ .
\ee
The set of the T-model potentials for $3\alpha = 1,2, 3,...,7$ is shown in Fig. \ref{Tmod}.
 In the leading approximation in 1/N, the predictions of E-models and T-models for the observational parameters $n_{s}$ and $r$ coincide with each other  for any given $\alpha$.  
However,
the value of  the inflaton field corresponding to the moment when the remaining number of e-foldings of inflation becomes equal to some number $N$ is slightly different, because of the different shape of the potential at small $\vp$ \cite{Kallosh:2016sej}:
\be\label{Tefold}
\vp_N\approx \sqrt {3\alpha \over 2} \log {8N \over 3 \alpha} \ .
\ee
For T-models with more general  potentials $V = m^{2}\tanh^{2n} {\vp\over \sqrt 6\alpha}$  one  has 
\be\label{Tefold2}
\vp_N\approx \sqrt {3\alpha \over 2} \log {8Nn \over 3 \alpha} \ .
\ee

\subsection{Two-disk models  and disc merger in half-plane variables}

In the two-disk case, we have just 4 moduli and would like to see the stabilization of 3 of them. By making some simple choices of the superpotential function $f(T)$ we can show that the relevant potentials acquire a plateau shape describing $\alpha$-attractors with $3\alpha = 1$ or $2$. A more complete dynamical picture will be revealed since we will be able to study not only the masses of the non-inflaton stabilized moduli near the inflaton trajectory, but the global properties of the models.

Here we start with
\be\label{K1}
K = -\frac12  \log   \frac{(T_1+\bar{T}_1)^2}{4T_1 \bar{T}_1}    - \frac12  \log   \frac{(T_2+\bar{T}_2)^2}{4T_2 \bar{T}_2}  + S\bar S\,.
\ee
Starting with a two-disk model, each a unit size one, we have two options: One can freeze dynamically one of the directions, e.g. $T_{2}$, by stabilizing it at some point $T_{2}  = c$, and get $3\alpha =1$ for inflation driven by the field $T_{1}$. Alternatively, one can enforce $T_{1} = T_{2}$ and get inflation with $3\alpha =2$.

\subsubsection{\boldmath $T_{2}=1$,  $3\alpha =1$} 
As an example of the model of two-disk model with $3\alpha =1$, we will study the theory with the superpotential
\be\label{w1}
W =  m  S (1-T_{1}) + M (1-T_2)^2 \,.     
\ee
Here $m \sim 10^{{-5}}$ is the inflaton mass scale, up to a factor $O(1)$, and $M \gg m$ is the stabilizing mass parameter. During inflation at $S=1-T_2=0$ supersymmetry is unbroken in the $T_1, T_2$ directions, but broken in the $S$  direction:
\ba
D_{T_1}W&=& - m S +  K_{T_1} \lp m S (1-T_{1}) + M (1-T_2)^2\rp\Big | _{S=1-T_2=0}=0\,,\cr
D_{T_2}W&=& 2M(1-T_2) + K_{T_2} \lp m S (1-T_{1}) + M (1-T_2)^2\rp\Big | _{S=1-T_2=0}=0\,,\cr
D_{S}W&=&  m  (1-T_{1}) +  \bar S  (\lp m S (1-T_{1}) + M (1-T_2)^2\rp\Big | _{S=1-T_2=0}= m (1-T_{1})\neq 0\,.\quad
\ea

\begin{figure}[H]
 \vspace*{3mm}
\centering
\includegraphics[width=9cm]{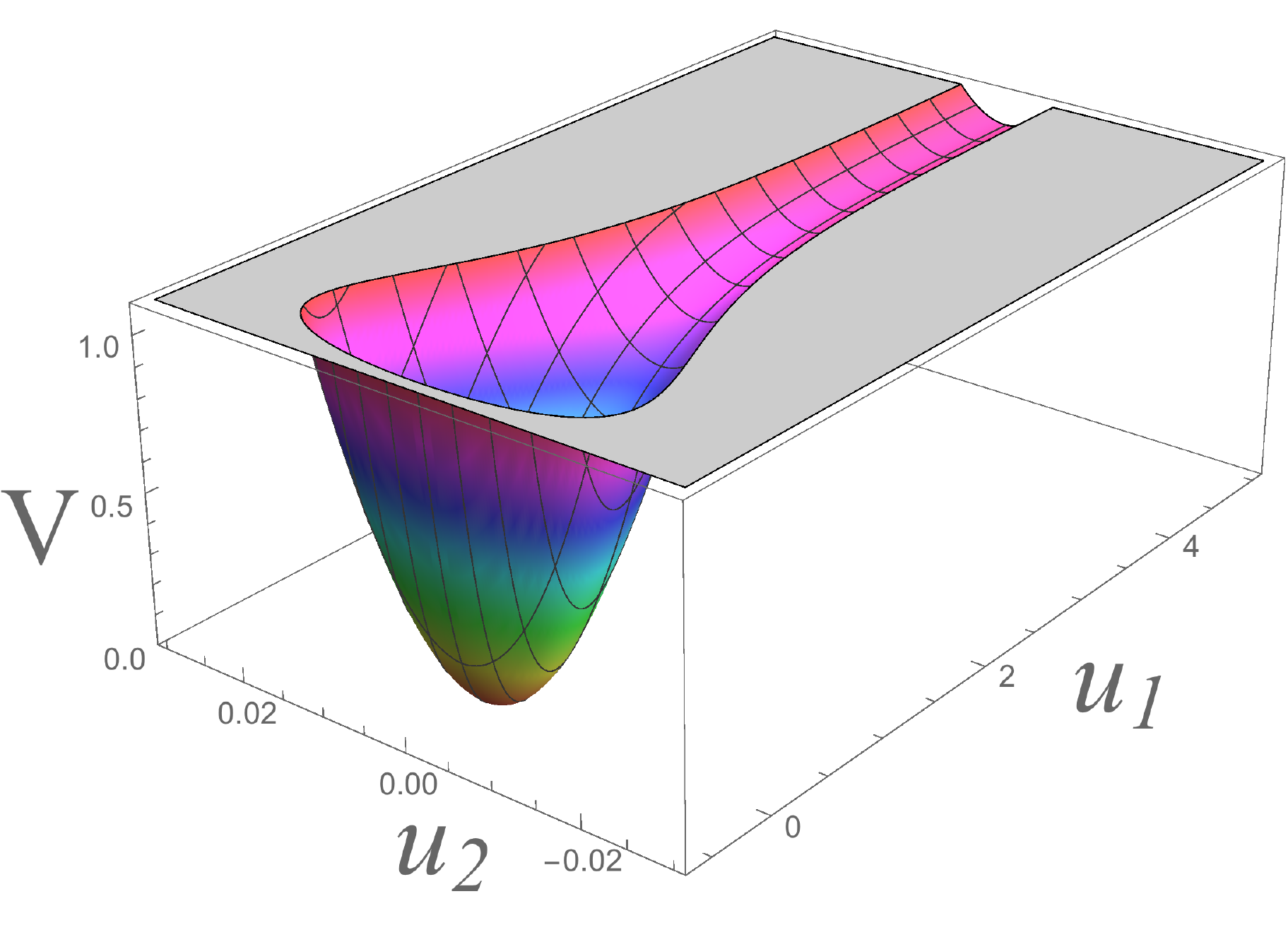}
\caption{\small Potential $V(u_{i})$ in the theory defined via equations \eqref{K1} and \rf{w1}. For $M = 10 m$ we have a heavy and stabilized $u_2$. The remaining field $u_{1}$ plays the role of the inflaton with the E-model $\alpha$-attractor potential  $V =m^{2}\bigl(1-e^{-\sqrt 2\, u_{1}}\bigr)^{2}$ corresponding to $3\alpha = 1$.  The fields are shown in Planckian units $M_{p}= 1$; the height of the potential is shown in units of $m^{2}$. }
\label{oo}
\end{figure}

As before, we will use the new variables $u_i$ and $a_{i}$, where  $T_i=e^{-\sqrt{2} u_i}(1+\rmi \sqrt{2} a_i )$. The variables   $u_i$ and $a_i$   become canonical in the limit of small $a_{i}$.    For $M = 0$ and $a_{1}=a_{2}= 0$, the potential depends only on the field $u_{1}$.   One can easily check that for any $u_{1}$ and  $M \not = 0$,  the fields $u_{2}$, $a_{1}$  and $a_{2}$ vanish  at the (local) minimum of the potential.  The potential of the field $u_{1}$ for  $u_{2}= a_{1}=a_{2}= 0$ is the standard potential \rf{EEE} of the E-model  $\alpha$-attractor with $3\alpha = 1$:
\be
V = m^{2}\left(1-e^{-\sqrt 2\, u_{1}}\right)^{2} \ .
\ee
The masses of the fields $a_{i}$ are always greater than the Hubble constant, so they are strongly stabilized at $a_{1}=a_{2}= 0$. The same is true for the field $u_{2}$, which is strongly stabilized at $u_{2}= a_{1}=a_{2}= 0$ for $M > {m\over 8\sqrt 3}$.   The inflationary potential in terms of $u_{1}$ and $u_{2}$ for $a_{i} = 0$ is shown in Fig. \ref{oo}. 

\subsubsection{\boldmath  A transition from  two $\alpha =1/3$ moduli to a single $\alpha =2/3$}\label{tr}
Now we study a model illustrating the dynamical merger of two $\alpha$-attractors with $\alpha =1/3$ to a single $\alpha$-attractor with $\alpha =2/3$.  We will consider the  superpotential 
\be
W  = m S \left(1-\frac{T_{1}+T_{2}}{2}\right) +M  (T_1-T_2)^2  \, .    
\ee
As we will see, the corresponding potential $V$ is very different from the one  studied in the previous subsection. 

We will investigate the potential $V$ using  variables $u_i$ and $a_{i}$, where  $T_i=e^{-\sqrt{2} u_i}(1+\rmi \sqrt{2} a_i )$.  One finds that the critical point equations $\partial_{a_i} V=0$ are solved for $a_{1}= a_{2} = 0$ and this solution corresponds to a minimum of the potential in these two directions. For large $M$, the two fields $u_{i}$ are merged during inflation into one canonically normalized field $\vp = (u_{1}+u_{2})/\sqrt 2 $ and the field orthogonal to it vanishes, $\chi =  (u_{1}-u_{2})/\sqrt 2 = 0$.  However, one can show that for $M \gg m$,  $\vp > {2\over 3}\log   {8\sqrt{2}M\over m}$ and $\chi = 0$ the field $\chi$ acquires a tachyonic mass, which leads to a tachyonic instability for the $\chi$ direction.

\begin{figure}[H]
\centering
\includegraphics[width=10cm]{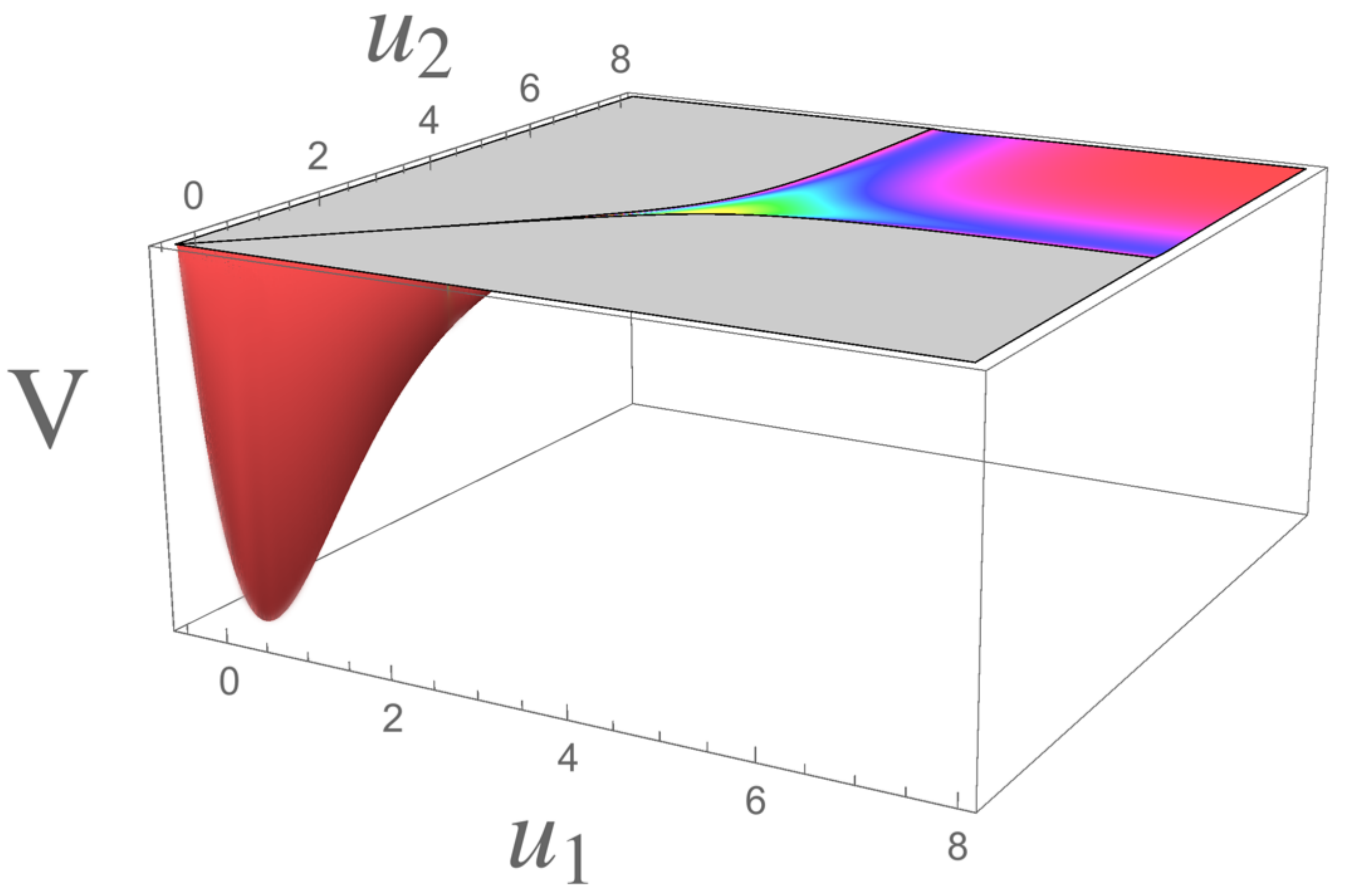}
\caption{\small The potential of the canonical fields $u_{i}$ for $M = 500 m$.  The height of the potential is shown in units $m^{2}$. The colored area shows the part of the potential with $V < m^{2}$; the red area in the upper right corner shows the inflationary plateau asymptotically approaching $V = m^{2}$. From there on, inflation continues when the field rolls towards the narrow gorge along which the fields $u_{1}$ and $u_{2}$ coincide. This gorge is seen as a narrow diagonal cut beginning at the center of the figure.  For  $\vp = (u_{1}+u_{2})/\sqrt 2 > {2\over 3}\log   {8\sqrt2M\over m}$, the gorge is wide, as seen in the right upper corner of the figure. The stable inflaton directions are shown by the blue valleys along which one of the $u_{i}$ remains nearly constant. The potential along each such direction asymptotically behaves as the $\alpha$-attractor potential with $3\alpha = 1$.  The merger of these two attractors, which occurs when  $\vp$ becomes smaller than  $ {2\over 3}\log   {8\sqrt2M\over m}$, corresponds to the phase transition to $3\alpha = 2$.}
\label{aa}
\end{figure}

\begin{figure}[h!]
\centering
\includegraphics[width=7cm]{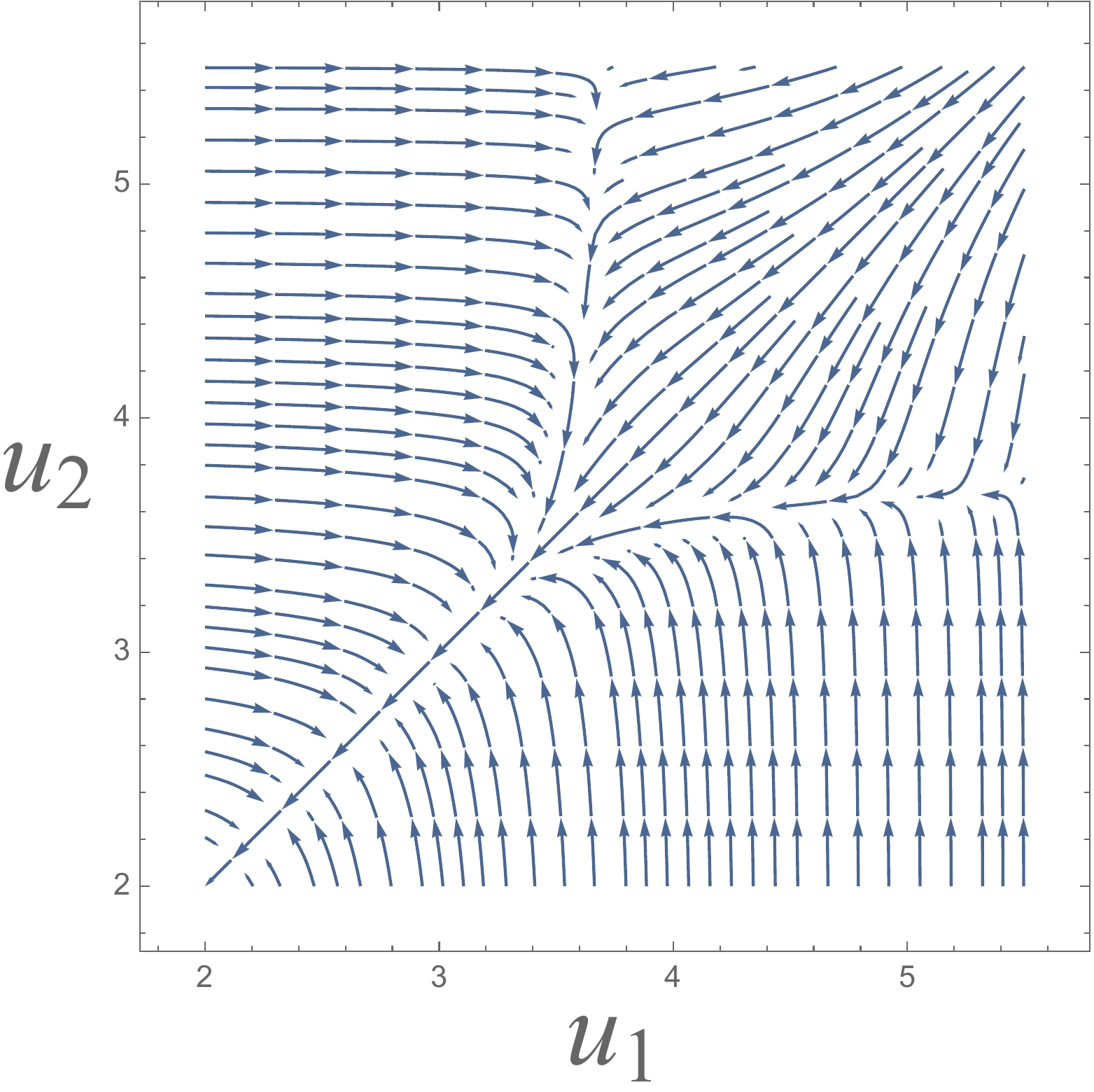}
\caption{\small Gradients of the inflationary potential. In the area where the slow-roll regime is possible (colored areas in Fig. \ref{aa}), these gradients describe the inflationary slow-roll evolution of the fields $u_{i}$. The fields starting their motion at the inflationary plateau in the right upper corner in Figs. \ref{aa} and Fig. \ref{bb} typically fall down towards one of the two streams, each of which can be  approximately  described as an $\alpha$-attractor  with $3\alpha = 1$. Then these streams merge into one stream corresponding to $3\alpha = 2$.}
\label{bb}
\end{figure}

The nature of this effect is illustrated by Figs. \ref{aa} and \ref{bb}. The colored area in Fig. \ref{aa} shows the part of the potential with $V < m^{2}$; the red area in the upper right corner shows an infinitely long inflationary plateau asymptotically approaching $V = m^{2}$. The fields tend to roll down from this plateau towards the narrow gorge, along which the fields $u_{1}$ and $u_{2}$ coincide. But at the early stages of this process the fields fall  towards one of the two stable inflaton directions shown by the two blue valleys in Fig. \ref{aa} along which one of the fields $u_{i}$ remains nearly constant, see the field flow diagram in Fig. \ref{bb}.

Note that this process is very slow to develop because the tachyonic mass of this field $\chi$ at the inflationary plateau with $M \gg m$ and  $\vp > {2\over 3}\log   {8\sqrt2M\over m}$ is exponentially small, much smaller than the Hubble constant,  just like the inflaton mass. Thus,  the field $\chi$, as well as the inflaton field, will experience inflationary fluctuations. But the general evolution of these fields  is dominated by their classical rolling, as shown in Fig. \ref{bb}.

The potential along each of the two  blue valleys in Fig. \ref{aa} asymptotically behaves as the $\alpha$-attractor potential with $\alpha = 1/3$.  When $\vp$ becomes smaller than $ {2\over 3}\log   {8 \sqrt2M\over m}$, the tachyonic mass of the field $\chi$ vanishes, this field becomes stable at $\chi = 0$, and the two inflaton directions merge into the inflationary gorge with  $\chi = 0$. This corresponds to symmetry restoration between $u_{1}$ and  $u_{2}$. The potential along the bottom of this gorge is  $V =m^{2}\left(1-e^{-\vp}\right)^{2}$,  which corresponds to $\alpha = 2/3$ in the E-model \rf{EEE}.   Quantum fluctuations of the field $\chi$ for a while remain large, until the positive mass squared of the field $\chi$ becomes greater than the Hubble constant squared $H^{2} \sim V/3$. This happens at  $\vp < {1\over 2} \log {16\sqrt 3 M\over m}$. This concludes the merger of the two inflaton directions and the phase transition from $\alpha = 1/3$ to $\alpha = 2/3$.

It is important to discuss the necessary condition that the effective trajectory $3\alpha=2$ lasts for $N$ e-foldings. The two axions have a positive mass $\mathcal{O}(H)$ even for large $\vp$, so we do not find any further constraints from their stabilization. However, as mentioned above, the $\chi$-direction may acquire a tachyonic mass for sufficiently large $\vp$, and there the inflaton trajectory bifurcates into two trajectories with $3\alpha=1$.  

Let us estimate how large $M$ is required to be in order to stabilize the field $\chi$ along the trajectory $3\alpha=2$ for the last $N$ e-foldings. According to \rf{efold}, the inflaton field value as a function of the e-folding number $N$ is given by $\vp_N\approx \sqrt {3\alpha \over 2} \log {4N \over 3 \alpha}$ for the  E-models we study.
Thus in our example with $3\alpha=2$ we find
$
\vp_N=\log(2N),\label{efold3}
$
and the mass of $\chi$ is
\begin{align}
m_{\chi}^2&=2e^{-4\vp}(128M^2-m^{2}e^{3\vp}+m^{2}e^{2\vp})=\frac{32M^2-2m^2N^3+m^2N^2}{2N^4}.\label{cmass}
\end{align}
From this expression, we obtain the following condition for $M$: 
\begin{align}\label{st1}
M > \frac{N^{3/2} m}{4} 
\end{align}
in the leading order in large $N$. For $N=55$, this constraint becomes $M \gtrsim 100 m$. To produce the observational result for the amplitude os scalar perturbations we need $m\sim 10^{-5}$ in Planck units, so $M$ should be greater than $\mathcal{O}(10^{-3})$. This implies that the inflationary trajectory $\chi = 0$, which corresponds to $\alpha$-attractor with $\alpha = 2/3$, becomes stable for $M \gtrsim 10^{{15}} - 10^{16}\text{GeV}$, which is well below the Planck scale and may correspond to the string/GUT scale, which is quite natural in our context.

It is also useful to know the condition that the mass of $\chi$ becomes larger than $H$, so that $\chi$ fluctuations are suppressed during the last $N$ e-folds. From Eq.~\eqref{cmass}, the condition $m_{\chi}>H$ can be satisfied for
\begin{align}\label{st2}
M>\frac{N^2}{4\sqrt{3}}m .
\end{align}
For $N=55$, this constraint reads $M \gtrsim 450 m$, and this can also be satisfied naturally, if $M$ is a Planck/string/GUT scale parameter.

Note that all the way until the field  $\vp$ becomes smaller than ${1\over 2} \log {16\sqrt 3 M\over m}$,  the mass squared of the field $\chi$ remains much smaller than $H$. In this regime, classical evolution of all fields typically brings them towards one of the two valleys corresponding to $\alpha = 1/3$, but details of this evolution may be somewhat affected by quantum fluctuations of the field $\chi$; see a discussion of a very similar regime in \cite{Demozzi:2010aj}. However, for $M>\frac{N^2}{4\sqrt{3}}m$ the evolution of the universe during the last $N$ e-foldings is described by the standard single field $\alpha$-attractor theory with $\alpha = 2/3$.

\subsection{Disk merger in disk variables}

In this section, we discuss the two-disk model in the disk variables $Z_i$ with
\be\label{DiskK2}
K = -\frac12 \sum_{i=1}^2 \log  \frac{(1-Z_i\bar Z_i)^2}{(1-Z_i^2)(1-\bar Z_i^2)} +S\bar S\,, \qquad 
W  = {m\over 2}\, S (Z_{1}+Z_{2}) +M  (Z_1-Z_2)^2  \, . 
\ee

\begin{figure}[H]
\begin{center}
\includegraphics[width=7.5cm]{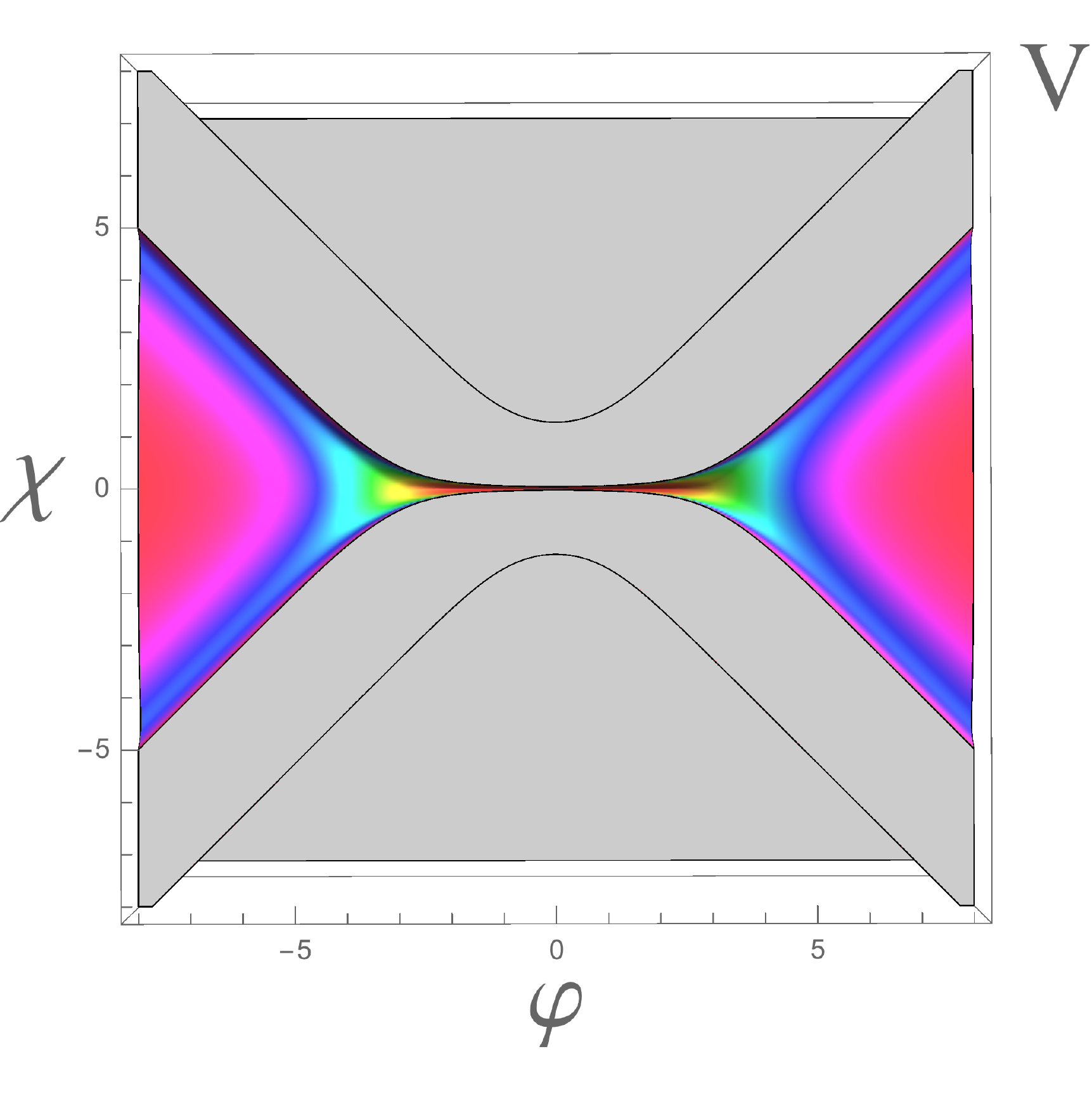}
\caption{\small A view from above on the inflationary potential. The red area shows the inflationary plateau asymptotically rising to $V = m^{2}$. The blue lines show valleys corresponding to inflationary attractors with $\alpha = 1/3$.  After the merger, they form a narrow gorge with the potential $V = m^{2} \tanh^{2} {\vp\over 2}$, which corresponds to the T-model $\alpha$-attractor with $\alpha = 2/3$ shown in Fig. \ref{ff}.   The value of the inflaton field $\vp$, which corresponds to the merger point, depends on the parameter $M$. For sufficiently large $M$, the last 60 e-folds of inflation are described by the single attractor with  $\alpha = 2/3$. }\label{ee}
\end{center}
\end{figure}

\begin{figure}[H]
\begin{center}
\includegraphics[width=8.5cm]{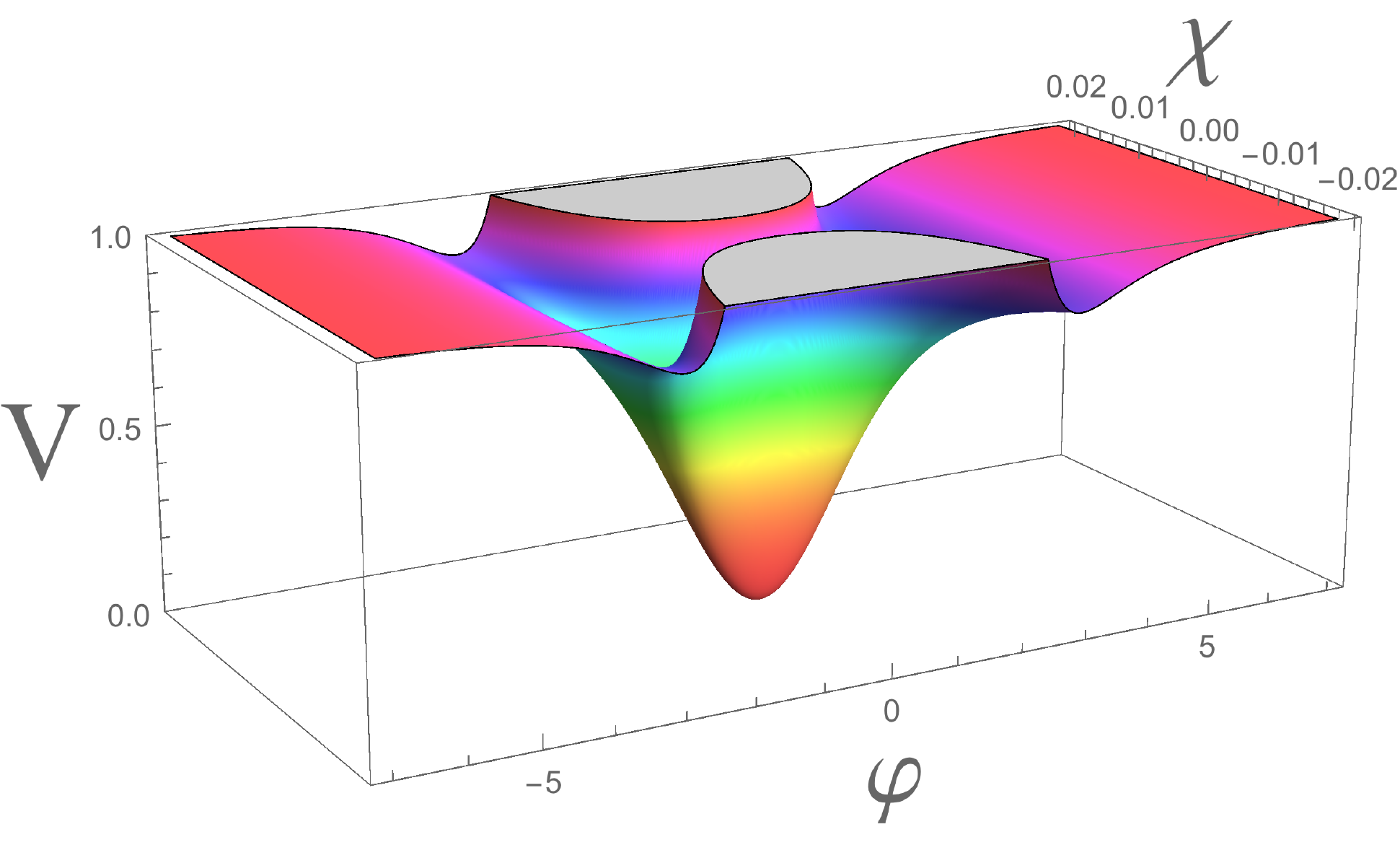}
\caption{\small A part of the Figure \ref{ee} shown in the same range of $\varphi$ as in Fig. \ref{ee},  but in a very narrow range of $\chi$ near the inflaton direction $\chi = 0$. One can easily recognize the T-model potential $V = m^{2} \tanh^{2} {\vp\over 2}$ with $3\alpha = 2$, which is  produced by merging of the   inflaton directions with $3\alpha = 1$ shown by the blue lines in the Figure \ref{ee}.  }
\label{ff}
\end{center} 
\end{figure}

As in the case with half-plane variables $T_i$, the stabilization of inflationary  trajectory at $Z_1=Z_2$ leads to the $\alpha$-attractor potential with $3\alpha=2$. We use the parametrization $Z_i=\tanh \frac{1}{\sqrt{2}}(\vp_i+\rmi\theta_i)$ such that $\vp_i$ and $\theta_i$ become canonical variables on the inflationary trajectory $\theta_i=0$. As in the E-model discussed in the previous section, for a large $M$, $\chi=\frac{1}{\sqrt2}(\vp_1-\vp_2)$ becomes heavy and is stabilized at the origin. There, the direction $\vp=\frac{1}{\sqrt2}(\vp_1+\vp_2)$ becomes the inflaton. Note that the axionic directions $\theta_i$ always have positive masses and are stabilized at $\theta_i=0$. The scalar potential in this T-model is given by  $V = m^{2} \tanh^{2} {\vp\over 2}$ and is shown in Figs.~\ref{ee} and \ref{ff}.


To make the merger last for $N$ e-foldings, the mass parameter $M$ should be sufficiently large,  as in the case of the E-model studied in the previous section. The axionic directions do not acquire tachyonic mass even for large $\vp$ but the field $\chi$  does. On the $3\alpha=2$ trajectory, the last e-folding number $N$ and the value of $\vp$ satisfy the same relation as in eq.~\eqref{Tefold}, and $m_{\chi}$ is given by
\begin{align}
m_{\chi}^{2}\approx\frac{16M^2-m^{2 }N^3}{N^4}
\end{align}
in the leading order in large $N$. 

The corresponding stability constraints on $M$ in this model are the same as the constraints \rf{st1}, \rf{st2} in the model studied in the previous section. The mass squared of the field $\chi$ is positive during the last $N$ e-foldings for $M> \frac{N^{3/2}}{4}\,m$. For $N=55$, this constraint becomes $M \gtrsim 100 m$.  The mass of the field $\chi$ becomes greater than the Hubble scale for $M> \frac{N^2}{4\sqrt3}\,m$.  For $N=55$, this is achieved for  $M \gtrsim 450 m$.  For $m \sim 10^{{-5}}$, the required value of the mass parameter $M$ should be greater than $10^{{-2}} - 10^{{-3}}$ in the Planck mass units.

\section{Back to the seven-disk scenario: Stability of the inflationary trajectory}

In the previous two sections we explored the effect of the merger of two disks, and found that for sufficiently strong stabilization described by the parameter $M$ one can easily stabilize the inflationary directory in such a way that during the last 50 - 60 e-foldings of inflation instead of two independent $\alpha$ attractors with $\alpha = 1/3$ one has a single inflaton potential with $\alpha = 2/3$. 

This investigation can be easily generalized for the 7-disk model studied in Section \ref{7d}, so here we will only present the result of this investigation of the theory in the half-plane variables, with the function  $f= m (1-\frac{1}{n}(T_{1}+\ldots +T_{n}))$, with $n \leq 7$.  Then, in analogy with equations \rf{st1}, \rf{st2}, one finds that at $N\gg 1$ the mass matrices of all fields orthogonal to the inflaton direction are positively definite, i.e. the inflationary trajectory is stable, for 
\be\label{7s}
M > {2 N^{3/2}m \over n^{{3}}} \ ,
\ee
and the strong stabilization with all of these masses  greater than the Hubble constant is achieved for
\be\label{7s2}
M > {2 N^{2}m\over   \sqrt 3 n^{{3}}} \ .
\ee
This result coincides with the result for the two-disk merger \rf{st1}, \rf{st2} for $n = 2$. It also shows that the merger of $n$ disks is easier to achieve for large $n$. In particular, for $n = 7$ the stability condition \rf{7s} during the last $N =55$ e-foldings of inflation is satisfied for   $M  \gtrsim  2.4 m$, and the strong stabilization condition \rf{7s2} of the $\alpha$-attractor regime with $\alpha = 7/3$ for  $N \lesssim 55$  is satisfied for $M  \gtrsim  10 m$.

For $\alpha =  7/3$, the Planck normalization for $m $ is  $m \approx \sqrt {7\over 3}\times  10^{{-5}}$, which leads to the stability condition  $M \gtrsim  7 \times  10^{{-5}}$ in the Planck mass units.

\section{Discussion}

In this paper we have used the relative simplicity of the general class of $\alpha$-attractor models,   \cite{Kallosh:2013hoa}-\cite{Galante:2014ifa}, to propose cosmological models with discrete values of the $\alpha$-parameter 
\be
3\alpha=R_{E}^2= 1, 2,\dots ,7\,.
\label{discrete1}\ee
These models realize the suggestion in \cite{Ferrara:2016fwe} that the consistent truncation of theories with maximal supersymmetry (M-theory, superstring theory, $\cN=8$ supergravity) to minimal $\cN=1$ supersymmetry models, leads to cosmological models with seven discrete values for the square of the radius of the Escher disk in moduli space.

It is instructive to remind us here that, if one would assume that the maximal supersymmetry models are first truncated to half-maximal supersymmetry models, for example, $\cN=4$ supergravity \cite{Cremmer:1977tt} and maximal $\cN=4$ superconformal models \cite{Bergshoeff:1980is}, one would recover the hyperbolic geometry with a single unit size disk. This would mean that $3\alpha=1$.

In the first part of the paper we have described the advantage of using the hyperbolic geometry of the moduli space to explain the relation between the tilt of the spectrum $n_s$ and the  number of e-foldings $N$, $n_{s} \approx 1-{2\over N}$, which is supported by the observational data. We also provided a geometric reason for $\alpha$-attractor models with plateau potentials, and explained the relation of hyperbolic geometry to the Escher's concept of `capturing infinity' in a finite space. Finally, we have shown that understanding `tessellations' of the hyperbolic disk, or equivalent of the half-plane geometry,  is useful for the  choice of the \K\, frame providing the stability of cosmological models. To derive the dynamical cosmological models supporting the case \rf{discrete1} we employ a two-step procedure:

As the first step,  in Sec. 5.1,  we introduce a  superpotential $W_0(T_i) $ in \rf{W0} where $T_i$ are coordinates of the seven-disk manifold.
It is consistent with $\cN=1$ supersymmetry, such that it has a supersymmetric minimum realizing dynamically the conditions on seven complex moduli \rf{all}, which were postulated in \cite{Ferrara:2016fwe}. The  superpotential depends on the parameter $m$, controlling the inflaton potential, and the parameter   $M$, which is responsible for the dynamical merger of the disks in a state where some of the moduli $T_{i}$ coincide. 

As a second step, in Sec. 5.2,  we introduce the cosmological sector of $\alpha$-attractor models. In addition to seven complex disk moduli we introduce a stabilizer superfield $S$ which can be either a nilpotent superfield, associated with the uplifting anti-D3 brane in string theory, or the one with  a very heavy scalar, which during inflation serves the purpose of stabilizing the non-inflaton directions. The corresponding \K\, potential  and superpotential are given in eqs. \eqref{eq:Kaehler} and \eqref{eq:Superpotential} respectively,  with $W_0$ given in eqn. \eqref{W0}. 

The original models  contain a rather large number of moduli: seven complex scalars and a stabilizer. We have studied these cosmological models close to the inflationary trajectory and we have found the conditions where the masses of all non-inflaton directions are positive. A more detailed study was performed in Sec. 6 in a toy model where the starting point is  a two-disk manifold, where the regimes with $3\alpha=1$ or $3\alpha=2$ are possible. We  studied these models  both in half-plane coordinates and in disk coordinates. The global analysis  of the cosmological evolution was performed, not just near the inflationary trajectory, and it was possible to evaluate the value of the parameter $M$  providing more than $N=55$ e-folds of inflation in the regime with $3\alpha=2$. Depending on the choice of $W_1$ we have found models with $M  \gtrsim  10^{-2}- 10^{-3} M_{Pl}$, 
describing inflation either with $3\alpha=1$, or $3\alpha=2$. A  cosmological phase transition between these two regimes is possible for smaller values of $M$. The analogous considerations for the seven-disk models in Sec. 7 show that the cosmological stability of the maximally symmetric regime with  $3\alpha=7$ requires $M  \gtrsim 10 m \sim  7 \times  10^{{-5}} M_{Pl}$.

Thus, in this paper we provided a dynamical realization of a new class of cosmological $\alpha$-attractors motivated by maximally supersymmetric theories, such as  M-theory, superstring theory,  and maximal $N = 8$  supergravity \cite{Ferrara:2016fwe}. These models suggest a set of discrete targets for the search of tensor modes in the range $10^{-3} \lesssim r \lesssim 10^{-2}$.  In particular, the maximally symmetric model $3\alpha=7$ and $r \approx 10^{-2}$ becomes an interesting realistic target for relatively early detection of B-modes. The case with $3\alpha=1$, $r\approx 10^{-3}$ remains a well motivated longer term goal.

For decades, one of the main goals of inflationary cosmology was to use observations to reconstruct the inflation  potential \cite{Lidsey:1995np}.  
From this perspective, it is especially interesting that in the new class of inflationary models, $\alpha$-attractors, the main cosmological predictions are determined  not by the potential, but by the hyperbolic geometry of the moduli space. This suggests that the cosmological observations probing the nature of inflation may tell us something important not only about the large-scale structure of our universe, but also about the geometry of the scalar manifold.

\section*{Acknowledgments}
We are grateful to  J. Carlstrom, S. Ferrara,   F. Finelli,  R. Flauger, J. Garcia-Bellido, S. Kachru,  C. L. Kuo,  L. Page, D. Roest,  E. Silverstein, F. Quevedo,   M. Zaldarriaga and A. Westphal for stimulating  discussions.
This work  is supported by SITP and by the NSF Grant PHY-1316699. The work of A.L. is also supported by the Templeton foundation grant  ``Inflation, the Multiverse, and Holography.''

\end{document}